\documentclass[12pt]{article}

\font\grande=cmr9.5 scaled \magstep4
\font\medio=cmr9.5 scaled \magstep2
\outer\def\beginsection#1\par{\medbreak\bigskip
      \message{#1}\leftline{\bf#1}\nobreak\medskip
\vskip-\parskip
      \noindent}
\usepackage{graphicx} 
\begin{document}
\bibliographystyle {unsrt}

\titlepage

\begin{flushright}
CERN-PH-TH/2013-224
\end{flushright}

\vspace{10mm}
\begin{center}
{\grande Inflationary susceptibilities, duality}\\
\vspace{10mm}
{\grande and large-scale magnetic fields generation}\\
\vspace{1.5cm}
 Massimo Giovannini
 \footnote{Electronic address: massimo.giovannini@cern.ch}\\
\vspace{1cm}
{{\sl Department of Physics, 
Theory Division, CERN, 1211 Geneva 23, Switzerland }}\\
\vspace{0.5cm}
{{\sl INFN, Section of Milan-Bicocca, 20126 Milan, Italy}}
\vspace*{0.5cm}
\end{center}

\vskip 0.5cm
\centerline{\medio  Abstract}
We investigate what can be said about the interaction of scalar fields with Abelian gauge fields during a quasi-de Sitter phase of expansion and under the assumption  that the electric and the magnetic susceptibilities do not coincide. The duality symmetry, transforming the magnetic susceptibility into the inverse of the electric susceptibility, exchanges the magnetic and electric power spectra. The mismatch between the two susceptibilities determines an effective refractive index affecting the evolution of the canonical fields. The constraints imposed by the duration of the inflationary phase and by the magnetogenesis requirements pin down the rate of variation of the susceptibilities that is consistent with the observations of the magnetic field strength over astrophysical and cosmological scales but avoids back-reaction problems. The parameter space of this magnetogenesis scenario is wider than in the case when the susceptibilities are equal, as it happens when the inflaton or some other spectator field is solely coupled to the standard gauge kinetic term.
\vskip 0.5cm

\noindent

\vspace{5mm}

\vfill
\newpage
\renewcommand{\theequation}{1.\arabic{equation}}
\setcounter{equation}{0}
\section{Introduction}
\label{sec1}
Large-scale magnetic field generation may take place in the early Universe \cite{rev1,rev2,rev3} and there are plausible reasons for this conjecture dubbed, some time ago, magnetogenesis \cite{mgenesis}. In this framework specific attention has been devoted to the interaction of gauge fields with scalar degrees of freedom during a quasi-de Sitter phase of expansion and in more general curved backgrounds relevant to cosmology (see, e.g. \cite{DT1,DT2,DT3,DT4,DT5,DT6,DT8,DT9} for a non exhaustive list of  references). The temperature and  
polarization anisotropies of the Cosmic Microwave Background (CMB in what follows) 
offer important clues about the origin of large-scale magnetism as repeatedly argued, along 
different perspectives (see e.g. \cite{rev1,rev2,rev3,mgenesis}), during the past score year.  A complete computation of the CMB observables  has been recently presented \cite{cmb} under the hypothesis that the same inflationary seed accounting for protogalactic magnetism also affects the EinsteinÐBoltzmann hierarchy whose initial conditions have been  directly bootstrapped out of the values provided by inflationary magnetogensis.

The aim of the present study is to discuss the idea that the electric and the magnetic susceptibilities may not coincide during inflation. So far such a possibility did not receive specific attention. For sake of definiteness consider the following action:
\begin{eqnarray}
S = - \frac{1}{16 \pi} \int \, d^{4} x\, \sqrt{-g} \biggl[ \lambda(\varphi,\psi) Y_{\alpha\beta} \, Y^{\alpha\beta} + {\mathcal M}_{\sigma}^{\rho}(\varphi) 
Y_{\rho\alpha}\, Y^{\sigma\alpha} - {\mathcal N}_{\sigma}^{\rho}(\psi) 
\tilde{Y}_{\rho\alpha}\, \tilde{Y}^{\sigma\alpha} \biggr],
\label{ac1}
\end{eqnarray} 
where $Y^{\mu\nu}$ and $\tilde{Y}^{\mu\nu}$ are, respectively, the gauge field strength and its dual;
 $g = \mathrm{det}g_{\mu\nu}$ is the determinant of the four-dimensional metric with signature 
 mostly minus. In the conventional case (see e.g. \cite{DT1,DT2,DT3,DT4,DT5,DT6,DT8,DT9})
${\mathcal M}_{\sigma}^{\rho}$ and ${\mathcal N}_{\sigma}^{\rho}$ are absent from Eq. (\ref{ac1}) so that the only coupling of the gauge fields to the scalar degrees of freedom is encoded in the first term inside the square bracket on the right hand side of Eq. (\ref{ac1}).  Suppose, as an example, that  ${\mathcal M}_{\sigma}^{\rho}$ and ${\mathcal N}_{\sigma}^{\rho}$ take the following form:
\begin{eqnarray}
 {\mathcal M}_{\sigma}^{\rho}(\varphi) &=& \frac{1}{2} \biggl(\partial_{\sigma} m_{E}^{*}\, \partial^{\rho} \,m_{E} + \partial_{\sigma} m_{E}\,\partial^{\rho} \,m_{E}^{*}\biggr),
 \nonumber\\
  {\mathcal N}_{\sigma}^{\rho}(\psi) &=&\frac{1}{2} \biggl(\partial_{\sigma} \,n_{B}^{*} \partial^{\rho}\, n_{B} + \partial_{\sigma} \,n_{B}\partial^{\rho} \,n_{B}^{*}\biggr),
\label{ac2}
\end{eqnarray}
where $m_{E} = m_{E}(\varphi)$ and $n_{B}= n_{B}(\psi)$. More complicated possibilities can 
be certainly imagined, like for instance $m_{E} = m_{E}(\varphi,\psi,\,...)$ and $n_{B}= n_{B}(\varphi,\psi,\, ...)$; the ellipses stand for other supplementary fields in case there are various inflatons or more than one spectator field \cite{lid1}. 
Equations (\ref{ac1})--(\ref{ac2}) describe the situation where the electric and the magnetic susceptibilities are not equal and include, as a special case, the
 following interaction
\begin{equation}
S = - \int d^{4} x \, \sqrt{-g}\,\biggl[ g_{1} \partial_{\alpha} \varphi \partial_{\beta} \varphi^{*} \, Y^{\alpha\rho} \, Y^{\beta}_{\, \rho} + g_{2} |\varphi|^2\, Y_{\alpha\beta} \, Y^{\alpha\beta} \biggr],
\label{ac3}
\end{equation}
that appears in the relativistic theory of Casimir-Polder and Van der Waals forces \cite{such}. Equation (\ref{ac3}) 
leads to static electric and magnetic susceptibilities that effectively depend on the scalar degrees of freedom; in the present study, the electric and the magnetic susceptibilities will be dynamical rather than static but still this analogy is physically instructive.  
Two further terms may arise in Eq. (\ref{ac1})
\begin{eqnarray}
S = - \frac{1}{16 \pi} \int \, d^{4} x\, \sqrt{-g} \biggl[ \overline{\lambda}(\psi)\,\, Y_{\alpha\beta} \, \tilde{Y}^{\alpha\beta} + \overline{{\mathcal M}}_{\sigma}^{\rho}(\psi) \,
Y_{\rho\alpha}\, \tilde{Y}^{\sigma\alpha}\biggr],
\label{acax}
\end{eqnarray} 
where $\psi$ may or may not coincide with the degrees of freedom mentioned above.
In the simplest case $\overline{{\mathcal M}}_{\sigma}^{\rho}=0$ and $\overline{\lambda} = \psi/M$: this is in a nutshell the coupling to the axions \cite{c0} which is not so effective for the amplification of gauge field fluctuations during a quasi-de Sitter stage of expansion \cite{c1,c2,c3}. The  pseudo-scalar vertex 
changes the topology of the magnetic flux lines once gauge field fluctuations have been already amplified \cite{mgt1}. For this reason the interactions appearing in Eq. (\ref{acax}) shall be neglected at least for the purposes of the 
present study.

If ${\mathcal M}_{\rho}^{\sigma}$ and ${\mathcal N}_{\rho}^{\sigma}$ are absent from Eq. (\ref{ac1}), the corresponding canonical Hamiltonian 
is explicitly invariant under electromagnetic duality \cite{duality1,duality3} when $\sqrt{\lambda} \to 1/\sqrt{\lambda}$. In practice this symmetry exchanges the magnetic and electric power spectra produced during a phase of quasi-de Sitter expansion and, more generally, in conformally flat backgrounds \cite{duality3}.
Whenever ${\mathcal M}_{\rho}^{\sigma}$ and ${\mathcal N}_{\rho}^{\sigma}$ are present a generalized duality symmetry transforms the magnetic susceptibility into the inverse of the electric susceptibility (and vice versa). 

The dynamical difference between electric and magnetic susceptibility affects the amplification of the quantum fluctuations 
of the gauge fields whose power spectra are related by duality. The computed power spectra can then be examined in the 
light of the magnetogenesis requirements and of other back-reaction constraints. The main purpose of this paper is not to endorse a specific set of initial conditions but to provide a comprehensive analysis of the whole idea. The parameter space of the model is wider than in the conventional case; both strongly and weakly coupled initial conditions are possible.

The paper is organized as follows. In  section \ref{sec2}, after some technical generalities,  we introduce the electric and the magnetic susceptibilities in conformally flat backgrounds and discuss the duality symmetry of the system. Section \ref{sec3} is devoted to the quantization of the problem and to 
the amplification of the quantum fluctuations of the gauge fields. The non-trivial evolution equations of  the  mode functions 
are solved in section \ref{sec4}; the power spectra are explicitly computed and related via the duality symmetry. 
Section \ref{sec5} contains the phenomenological considerations 
related to magnetogenesis. The concluding remarks are collected in section \ref{sec6}.

\renewcommand{\theequation}{2.\arabic{equation}}
\setcounter{equation}{0}
\section{Generalities}
\label{sec2}
\subsection{Preliminary considerations}
From the action (\ref{ac1}) the following equations of motion can be easily derived:
\begin{eqnarray}
&& \nabla_{\alpha} \biggl( \lambda \, Y^{\alpha\beta}\biggr) + \frac{1}{2} \nabla_{\alpha} {\mathcal Z}^{\alpha\beta}  - \frac{1}{2} \nabla_{\alpha} {\mathcal W}^{\alpha\beta} = 4 \pi j^{\beta},
\label{FF1}\\
&& \nabla_{\alpha} \tilde{Y}^{\alpha\beta} =0, 
\label{FF2}
\end{eqnarray}
where $\nabla_{\alpha}$ is the covariant derivative; the two antisymmetric tensors 
${\mathcal Z}^{\alpha\beta}$ and ${\mathcal W}^{\alpha\beta}$ are:
\begin{eqnarray}
{\mathcal Z}^{\alpha\beta} &=& {\mathcal M}^{\alpha}_{\sigma} \,\,Y^{\sigma\beta} - {\mathcal M}^{\beta}_{\sigma} \,\, Y^{\sigma\alpha},
\label{Z}\\
{\mathcal W}^{\alpha\beta} &=& E^{\alpha\beta\rho\zeta} \,\, \tilde{Y}_{\sigma\zeta}\, \, {\mathcal N}^{\sigma}_{\rho}
\label{W}\\
&=& {\mathcal N}^{\beta}_{\,\rho} \, Y^{\alpha\rho} -  {\mathcal N}^{\alpha}_{\,\rho} \, Y^{\beta\rho} - {\mathcal N}^{\rho}_{\,\rho} \, Y^{\alpha\beta}. 
\label{W1}
\end{eqnarray}
In Eq. (\ref{W}) $E^{\alpha\beta\rho\zeta}= \epsilon^{\alpha\beta\rho\zeta}/\sqrt{-g}$ and $\epsilon^{\alpha\beta\rho\zeta}$ is the total antisymmetric pseudotensor of fourth rank. Equations (\ref{FF1}) and (\ref{FF2}) can also be recast in the following form:
\begin{eqnarray}
&& \frac{1}{\sqrt{- g}} \partial_{\alpha} \biggl[ \sqrt{-g} \lambda Y^{\alpha\beta}\biggr] +  \frac{1}{2 \sqrt{- g}}\partial_{\alpha}\biggl[ \sqrt{-g}\, {\mathcal Z}^{\alpha\beta} \biggr]  -  \frac{1}{2 \sqrt{- g}}\partial_{\alpha}\biggl[ \sqrt{-g}\, {\mathcal W}^{\alpha\beta}\biggr] = 4 \pi \, j^{\beta},
\label{V1}\\
&& \frac{1}{\sqrt{-g}} \partial_{\alpha} \biggl[ \sqrt{-g} \, \tilde{Y}^{\alpha\beta} \biggr] =0.
\label{V2}
\end{eqnarray}
The tensors  ${\mathcal M}^{\rho}_{\sigma}$ and of ${\mathcal N}^{\rho}_{\sigma}$  shall now be parametrized as: 
\begin{equation}
{\mathcal M}^{\rho}_{\sigma}(\varphi) = \lambda_{E}(\varphi) \, u^{\rho} \, u_{\sigma}, \qquad {\mathcal N}^{\rho}_{\sigma}(\psi)= \lambda_{B}(\psi) \, \overline{u}^{\rho} \, \overline{u}_{\sigma},
\label{V4}
\end{equation}
where 
\begin{equation}
u_{\rho} = \frac{\partial_{\rho}\varphi}{\sqrt{g^{\alpha\beta} \partial_{\alpha} \varphi \partial_{\beta} \varphi}}, \qquad 
\overline{u}_{\rho} = \frac{\partial_{\rho}\psi}{\sqrt{g^{\alpha\beta} \partial_{\alpha} \psi \partial_{\beta} \psi}},
\end{equation}
and $g^{\alpha\beta} \, u_{\alpha} \, u_{\beta} = 1$,  $g^{\alpha\beta} \, \overline{u}_{\alpha} \, \overline{u}_{\beta} = 1$. The validity of the parametrization (\ref{V4}) can be verified by inserting, for instance\footnote{Note that $\theta_{\varphi}$ and $\theta_{\psi}$ are dimensionless 
constants while $M_{\varphi}$ and $M_{\psi}$ are two different mass scales.}, 
$m_{E}(\varphi) = \exp{(\theta_{\varphi} \varphi/M_{\varphi})}$ and $n_{B}(\psi) = \exp{(\theta_{\psi} \psi/M_{\psi})}$
into Eq. (\ref{ac2}). 
Equation (\ref{V4}) implies that $u_{\mu}$ and $\overline{u}_{\mu}$ are invariant under the reparametrizations of $\varphi$ and $\psi$, i.e. $\varphi\to \Phi = q_{1}(\varphi)$ and $\psi \to \Psi = q_{2}(\psi)$. For these reasons different models may lead to the same $\lambda_{E}$ and $\lambda_{B}$ and 
the parametrization of Eq. (\ref{V4}) is sufficiently general for the present ends.
Using Eq. (\ref{V4}) the action (\ref{ac1}) can be recast in the following form:
\begin{equation}
S = - \frac{1}{16 \pi} \int d^{4} x \sqrt{ -g} \biggl[ \lambda \,Y_{\alpha\beta} \, Y^{\alpha\beta} + \lambda_{E} \,u^{\rho} \, u_{\sigma} \, Y_{\rho\alpha} \, Y^{\sigma\alpha}- \lambda_{B} \, \overline{u}^{\rho} \,\overline{u}_{\sigma} \,\tilde{Y}_{\rho\alpha} \, \tilde{Y}^{\sigma\alpha} \biggr].
\label{ac1A}
\end{equation}
Equation (\ref{ac1A}) elucidates the connection of  $\lambda_{E}$ and $\lambda_{B}$ with the electric and magnetic susceptibilities. In fact $u_{\rho} \tilde{Y}^{\alpha\rho} = {\mathcal B}^{\alpha}$ and $u_{\rho} Y^{\alpha\rho} = {\mathcal E}^{\alpha}$ are the electric and magnetic fields in covariant form as it follows from the generally covariant decomposition of the gauge field strengths \cite{lic}:
\begin{eqnarray}
Y_{\alpha\beta} &=& {\mathcal E}_{\alpha} u_{\beta} - {\mathcal E}_{\beta} u_{\alpha} + E_{\alpha\beta\rho\sigma} \, u^{\rho} \, {\mathcal B}^{\sigma},
\nonumber\\
\tilde{Y}^{\alpha\beta} &=& {\mathcal B}^{\alpha} u^{\beta} - {\mathcal B}^{\beta} u^{\alpha} + E^{\alpha\beta\rho\sigma} \, {\mathcal E}_{\rho} \, u_{\sigma},
\end{eqnarray}
where the four-velocity may coincide either with $u_{\rho}$ or with $\overline{u}_{\rho}$.
The functional ${\mathcal M}_{\rho}^{\sigma}$ and ${\mathcal N}_{\rho}^{\sigma}$ can be split into a homogeneous 
part and an inmohomogeous part, i.e. 
\begin{eqnarray}
{\mathcal M}_{0}^{0}(\tau) &=& \lambda_{E}(\tau) \, u_{0}\, u^{0}, \qquad {\mathcal N}_{0}^{0}(\tau) = \lambda_{B}(\tau) \, \overline{u}_{0}\, \overline{u}^{0},  
\nonumber\\
{\mathcal M}_{i}^{0}(\vec{x},\tau) &=& \lambda^{(1)}_{E}(\vec{x},\tau)\, u_{i}\, u^{0},\qquad {\mathcal N}_{i}^{0}(\vec{x},\tau) = \lambda^{(1)}_{B}(\vec{x},\tau)\, \overline{u}_{i}\, \overline{u}^{0},
\nonumber\\
{\mathcal M}_{i}^{j}(\vec{x},\tau) &=& \lambda^{(1)}_{E}(\vec{x},\tau)\, u_{i}\, u^{j}, \qquad 
\qquad {\mathcal N}_{i}^{j}(\vec{x},\tau) = \lambda^{(1)}_{B}(\vec{x},\tau)\, \overline{u}_{i}\, \overline{u}^{j},
\label{MMNN} 
\end{eqnarray}
where $\lambda_{E}(\vec{x},\tau) = \lambda_{E}(\tau) + \lambda^{(1)}_{E}(\vec{x},\tau)$ and   $\lambda_{B}(\vec{x},\tau) = \lambda_{B}(\tau) + \lambda^{(1)}_{B}(\vec{x},\tau)$. The various contributions can be taken into account, order by order, within the 
standard perturbative expansion involving the fluctuations of the scalar degrees of freedom and of the geometry\footnote{The leading order contribution of  Eq. (\ref{MMNN}), i.e. the fully homogeneous part, is, in a sense, more general than the original action insofar as it can even parametrize the case where the interaction is not in the form of Eq. (\ref{ac2}); some examples along this direction are  $\partial^{\rho}\partial_{\sigma} \varphi Y_{\rho\alpha} Y^{\sigma\alpha}$ or $\partial^{\rho}\partial_{\sigma} \psi \tilde{Y}_{\rho\alpha} \tilde{Y}^{\sigma\alpha}$.}.
Finally, from the action (\ref{ac1}) the energy-momentum tensor of the gauge fields reads:
\begin{equation}
T_{\mu}^{\nu} = \frac{1}{4\pi}\, \biggl[ - {\mathcal S}_{\mu}^{\nu} + \frac{1}{4} \, {\mathcal S} \, \delta_{\mu}^{\nu}\biggr],
\label{ENM1}
\end{equation}
where 
\begin{eqnarray}
{\mathcal S}_{\mu}^{\nu} &=& \lambda Y_{\alpha\mu} \, Y^{\alpha\nu} + \frac{1}{2}\biggl( {\mathcal M}^{\rho}_{\mu} \, Y_{\rho\alpha} \, Y^{\nu\alpha} + {\mathcal M}^{\rho}_{\sigma} \, Y_{\rho\mu} \,Y^{\sigma\nu} \biggr)
\nonumber\\
&-& \frac{1}{2} \biggl( {\mathcal N}^{\rho}_{\mu} \, \tilde{Y}_{\rho\alpha} \, \tilde{Y}^{\nu\alpha} + {\mathcal N}^{\rho}_{\sigma} \tilde{Y}_{\rho\mu}\, \tilde{Y}^{\sigma\nu} \biggr).
\label{ENM2}
\end{eqnarray}
\subsection{Conformally flat backgrounds}
Consider the case of a conformally flat 
metric $g_{\mu\nu} = a^2(\tau) \, \eta_{\mu\nu}$ where $a(\tau)$ is the scale factor and $\eta_{\mu\nu}$ is the Minkowski metric. To lowest order ${\mathcal M}_{0}^{0}(\tau) = \lambda_{E}(\tau)$ and ${\mathcal N}_{0}^{0}(\tau) = \lambda_{B}(\tau)$ are homogeneous (see Eq. (\ref{MMNN})) while all the other entries 
are inhomogeneous. Recalling that  $Y^{i 0} = e^{i}/a^2$  and $Y^{ij} = - \epsilon^{ijk} b_{k}/a^2$ (where $\vec{e}$ and $\vec{b}$ are the electric and magnetic fields in flat space-time), Eqs. (\ref{FF1})--(\ref{FF2}) and (\ref{V1})--(\ref{V2}) can be written as:
\begin{eqnarray}
&& \vec{\nabla} \cdot \biggl[ a^2 \biggl( \lambda + \frac{\lambda_{E}}{2} \biggr) \vec{e} \biggr] = 4 \pi \rho,
\nonumber\\
&& \vec{\nabla}\times \biggl[ a^2 \biggl( \lambda + \frac{\lambda_{B}}{2} \biggr) \vec{b} \biggr] 
= \partial_{\tau}  \biggl[ a^2 \biggl( \lambda + \frac{\lambda_{E}}{2} \biggr) \vec{e} \biggr] + 4 \pi \vec{J}, 
\nonumber\\
&& \vec{\nabla} \cdot ( a^2 \vec{b}) =0, \qquad \partial_{\tau} ( a^2 \vec{b}) + \vec{\nabla} \times ( a^2 \vec{e}) =0,
\label{V5}
\end{eqnarray}
where $\rho$ and $\vec{J}$ denote the electromagnetic sources that are important both at the beginning and at the end of the inflationary evolution\footnote{During the protoinflationary stage of expansion, the electromagnetic sources are not immediately washed out because there exist symmetries preventing their dissipation \cite{DT9}. At the end of inflation charged particles must be included as they  determine the effective post-inflationary conductivity. The total charge density vanishes on its own since the initial plasma, even if present, must be globally neutral.}. Introducing the following rescaled fields
\begin{eqnarray}
\vec{B} &=& a^2 \, \sqrt{\Lambda_{B}}\, \vec{b}, \qquad \vec{E} = a^2 \, \sqrt{\Lambda_{E}}\, \vec{e},
\label{defEB}\\
\Lambda_{B} &=& \lambda + \frac{\lambda_{B}}{2}, \qquad \Lambda_{E} = \lambda + \frac{\lambda_{E}}{2},
\end{eqnarray}
the system of Eq. (\ref{V5}) becomes:
\begin{eqnarray}
&& \vec{\nabla} \times \biggl( \sqrt{\Lambda_{B}} \vec{B} \biggr) = \partial_{\tau} \biggl( \sqrt{\Lambda_{E}} \vec{E} \biggr) + 4 \pi \vec{J},
\label{first}\\
&& \vec{\nabla} \times \biggl(\frac{\vec{E}}{\sqrt{\Lambda_{E}}}\biggr) + \partial_{\tau} \biggl(\frac{\vec{B}}{\sqrt{\Lambda_{B}}}\biggr) =0,
\label{second}\\
&& \vec{\nabla} \cdot \biggl(\frac{\vec{B}}{\sqrt{\Lambda_{B}}}\biggr)=0,\qquad \vec{\nabla}\cdot ( \sqrt{\Lambda_{E}}\, \vec{E} ) = 4 \pi \rho.
\label{third}
\end{eqnarray}
The electric and the magnetic susceptibilities $\chi_{E}$ and $\chi_{B}$ are defined as:
\begin{equation}
\chi_{E} = \sqrt{\Lambda_{E}} \equiv \sqrt{ f\, \Lambda_{B}}, \qquad \chi_{B} = \sqrt{\Lambda_{B}} \equiv\sqrt{ \frac{\Lambda_{E}}{f}}, \qquad f= \biggl(\frac{\chi_{E}}{\chi_{B}}\biggr)^2 = \frac{\Lambda_{E}}{\Lambda_{B}}.
\label{chi}
\end{equation}
From the Eqs. (\ref{ENM1})--(\ref{ENM2}) it is possible to deduce the various components of the energy-momentum tensor 
by recalling the relation of the gauge field strengths to the physical fields. Consider, for instance, the energy density 
always in the case of the general parametrization discussed above
\begin{equation}
T_{0}^{0} =  \frac{1}{8 \pi}\biggl[ \chi_{E} e^2 + \chi_{B} b^2\biggr],
\label{enm3}
\end{equation}
which can be expressed in terms of the rescaled fields $\vec{E}$ and $\vec{B}$, when needed. Similar 
manipulations can be used to deduce the other components  of the energy-momentum tensor.
\subsection{Duality properties}
Neglecting the sources, Eqs. (\ref{first}), (\ref{second}) and (\ref{third}) are invariant under the following set of transformations:
 \begin{equation}
 \vec{E} \to - \vec{B}, \qquad \vec{B} \to \vec{E}, \qquad \chi_{B} \to 
 \frac{1}{\chi_{E}},\qquad \chi_{E}  \to 
 \frac{1}{\chi_{B}},
 \label{dual3}
 \end{equation}
leaving unaltered the ratio $f= (\chi_{E}/\chi_{B})^2$. The duality properties are manifest from the decoupled evolution of the electric and of the magnetic fields:
\begin{eqnarray}
&& \frac{1}{\chi_{B}}\partial_{\tau} \biggl[ \chi_{E}^2 \partial_{\tau} \biggl(\frac{\vec{B}}{\chi_{B}}\biggr)\biggr] - \nabla^2 \vec{B} 
= \frac{4\pi}{\chi_{B}} \vec{\nabla} \times \vec{J},
\label{dec1}\\
&& \chi_{E} \,\partial_{\tau} \biggl[ \frac{1}{\chi_{B}^2}\partial_{\tau} \biggl(  \chi_{E}\, \vec{E} \biggr) + \frac{4\pi}{\chi_{B}^2} \vec{J} 
\biggr] - \nabla^2 \vec{E} =0.
\label{dec2}
\end{eqnarray}
Equations (\ref{dec1}) and (\ref{dec2}) can be written, more explicitly, as:
\begin{eqnarray}
&& \vec{E}'' + 2 \biggl(\frac{\chi_{E}}{\chi_{B}}\biggr)^{\prime} \biggl(\frac{\chi_{B}}{\chi_{E}}\biggr)  \vec{E}' + \biggl[ 
\frac{\chi_{E}''}{\chi_{E}} - 2 \biggl( \frac{\chi_{E}'}{\chi_{E}}\biggr)\biggr( \frac{\chi_{B}'}{\chi_{B}}\biggr)\biggr] \vec{E} - \frac{\nabla^2 \vec{E}}{f} = - 4\pi \frac{\chi_{E}}{f} \biggl(\frac{\vec{J}}{\chi_{B}}\biggr)^{\,\prime},
\label{dec1a}\\
&& \vec{B}'' + 2 \biggl(\frac{\chi_{E}}{\chi_{B}}\biggr)^{\prime} \biggl(\frac{\chi_{B}}{\chi_{E}}\biggr)\vec{B}' + \biggl[ 
\chi_{B} \biggl(\frac{1}{\chi_{B}}\biggr)^{\prime\prime} - 2 \biggl( \frac{\chi_{E}'}{\chi_{E}}\biggr)\biggr( \frac{\chi_{B}'}{\chi_{B}}\biggr)\biggr] \vec{B} - \frac{\nabla^2 \vec{B}}{f} = \frac{4\pi \vec{\nabla} \times \vec{J}}{\chi_{B} f},
\label{dec2a}
\end{eqnarray}
where the prime denotes a derivation with respect to the conformal time coordinate $\tau$.
Recalling Eq. (\ref{dual3}), the system of Eqs. (\ref{dec1a}) and (\ref{dec2a}) is left invariant 
by Eq. (\ref{dual3}) provided $\vec{J} =0$.  Using a further rescaling of the electric and magnetic fields (i.e.  $\vec{Q}_{B} = \sqrt{f} \, \vec{B}$ and  $\vec{Q}_{E} = \sqrt{f} \, \vec{E}$), Eqs. (\ref{dec1a}) and (\ref{dec2a}) can be simplified by eliminating the first time derivatives:
\begin{eqnarray}
&& \vec{Q}_{B}^{\prime\prime} \, - \frac{\nabla^2 \vec{Q}_{B}}{f} \, - \frac{(\chi_{B} \sqrt{ f})''}{\chi_{B} \sqrt{f}}\,  \vec{Q}_{B} = \frac{4\pi \,\vec{\nabla} \times \vec{J}}{ \chi_{B} \sqrt{f}},
\label{QB}\\
&&  \vec{Q}_{E}^{\prime\prime} \, - \frac{\nabla^2 \vec{Q}_{E}}{f}\, - \, \biggl(\frac{\sqrt{f}}{\chi_{E}}\biggr)'' \biggl(\frac{\chi_{E}}{f}\biggr) \,\vec{Q}_{E} = \, - \frac{4 \pi \,\chi_{E}}{\sqrt{f}}\, 
\biggl(\frac{\vec{J}}{\chi_{B}^2} \biggr)'.
\label{QE}
\end{eqnarray}
  From Eq. (\ref{defEB}) we have that 
$\vec{B} = \, a^2 \, \vec{\nabla} \times( \chi_{B} \vec{Y})$ and $\vec{E} = -\chi_{E} \, \partial_{\tau} \vec{Y}$ where 
$\vec{Y}$ is the vector potential in the gauge $Y_{0} =$ and $\vec{\nabla}\cdot \vec{Y} =0$. 
As specifically discussed in the section \ref{sec3}, the canonical normal mode of the action is related to the vector potential $\vec{Y}$ as $\vec{Y} = \vec{A}/\chi_{E}$ up to a constant that depends on the system of units. Consequently the relation of the electric and magnetic fields to the canonical vector potential is:
\begin{eqnarray}
&& \vec{B} =  \vec{\nabla} \times\biggl( \frac{\chi_{B}}{\chi_{E}} \vec{A} \biggr)= \vec{\nabla} \times 
\biggl(\frac{\vec{A}}{\sqrt{f}}\biggr),
\label{normB}\\
&&\vec{E} = - \chi_{E}\, \biggl(\frac{\vec{A}}{\chi_{E}} \biggr)^{\,\prime} = - \vec{A}^{\,\,\prime} + \frac{\chi_{E}^{\,\,\prime}}{\chi_{E}} \vec{A}.
\label{normE}
\end{eqnarray}
Inserting Eqs. (\ref{normB}) and  (\ref{normE}) into Eq. (\ref{first}) the equation obeyed by $\vec{A}$ can be obtained and solved; this analysis will be postponed to
 sections \ref{sec3} and \ref{sec4}. It is finally useful to discuss, in some detail, the limit $\chi_{E} \to \chi_{B}$ (or, which is the same, $f \to 1$). 
When  $f\to 1$ the following relations can be explicitly verified:
\begin{equation}
\lim_{f\to 1} \chi_{B} = \lim_{f\to 1} \chi_{E} = \sqrt{\lambda},\qquad \lim_{f\to 1} \vec{Q}_{B} = \vec{B}, \qquad 
\lim_{f\to 1} \vec{Q}_{E} = \vec{E}.
\label{limit1}
\end{equation}
Using Eq. (\ref{limit1}) into Eqs. (\ref{QB}) and (\ref{QE}) we obtain
\begin{eqnarray}
&& \vec{B}^{\prime\prime} \, - \nabla^2 \vec{B} \, - \frac{\sqrt{\lambda}^{\,\prime\prime}}{\sqrt{\lambda}}\,  \vec{B} = 
\frac{4\pi \vec{\nabla} \times \vec{J}}{ \sqrt{\lambda} } ,
\label{B1}\\
&&  \vec{E}^{\prime\prime} \, - \nabla^2 \vec{E}\, - \, \biggl(\frac{1}{\sqrt{\lambda}}\biggr)^{\prime\prime} \sqrt{\lambda} \,\vec{E} = \, - 4 \pi \sqrt{\lambda} \, 
\biggl(\frac{\vec{J}}{\lambda} \biggr)',
\label{E1}
\end{eqnarray}
which is the standard result obtainable in the case when ${\mathcal M}_{\rho}^{\sigma}\to 0$ and ${\mathcal N}_{\rho}^{\sigma}\to 0$ in Eq. (\ref{ac1}) (see, for instance, \cite{DT9,duality3}).

\renewcommand{\theequation}{3.\arabic{equation}}
\setcounter{equation}{0}
\section{Quantum fluctuations}
\label{sec3}
\subsection{Canonical Hamiltonian}
Consider Eqs. (\ref{ac1}) and (\ref{ac1A}), in time-dependent (conformally flat) backgrounds and in the Coulomb gauge (i.e. $Y_{0} =0$  and $\vec{\nabla}\cdot \vec{Y} =0$)  that is preserved (unlike the Lorentz gauge condition)  under a conformal rescaling of the metric. The action (\ref{ac1A}) becomes:
\begin{eqnarray}
S &=& \int \,d\tau\, L(\tau), \qquad L(\tau) = \int d^{3} x\, {\mathcal L}(\vec{x},\tau),
\label{dual5}\\
{\mathcal L}(\vec{x},\tau) &=& \frac{1}{2} \biggl\{ \vec{A}^{\,\prime \,2} + \biggl(\frac{\chi_{E}^{\,\prime}}{\chi_{E}}\biggr)^2 
 \vec{A}^{\,2}  - 2 \frac{\chi_{E}'}{\chi_{E}} \vec{A} \cdot \vec{A}^{\,\prime} - \frac{\chi_{B}^2}{\chi_{E}^2}\partial_{i} \vec{A} \cdot \partial^{i} \vec{A}\biggr\},
\label{dual6}
\end{eqnarray}
where\footnote{The $1/\sqrt{4\pi}$ is purely conventional 
and its presence comes from the factor $16\pi$ included in the initial gauge action of Eq. (\ref{ac1}). } $\vec{A} = \sqrt{ \Lambda_{E}/(4\pi)} \vec{Y}$. We have assumed that $\chi_{E}$ and $\chi_{B}$ are only dependent on the conformal time coordinate $\tau$. The canonical momentum conjugate to $\vec{A}$ is obtained from Eq. (\ref{dual6}) and it coincides, up to a sign, with the canonical electric field, i.e. 
\begin{equation}
\vec{\pi} = \vec{A}^{\,\prime} - \frac{\chi_{E}'}{\chi_{E}} \vec{A} = - \vec{E},
\label{dual7}
\end{equation}
while, as already discussed, $\vec{B} =  \vec{\nabla} \times (\vec{A}/\sqrt{f})$. The canonical Hamiltonian is then given by 
\begin{equation}
H_{A}(\tau) = \frac{1}{2} \int d^3 x \biggl[ \vec{\pi}^{2} + 2 \frac{\chi_{E}'}{\chi_{E}} \vec{\pi} \cdot \vec{A} + 
 \frac{\partial_{i} \vec{A} \cdot \partial^{i} \vec{A}}{f}\biggr].
\label{dual8}
\end{equation}
Since $\chi_{E}$, $\chi_{B}$ and $f$ are not three independent functions, only two of them can be independently assigned.  It is practical to select $\chi_{E}$ and $f$ independently while $\chi_{B}$ 
can be derived as $\chi_{B} = \chi_{E}/\sqrt{f}$.  The Fourier mode expansion for the canonical fields reads
\begin{equation}
\vec{\pi}(\vec{x},\tau) = \frac{1}{(2\pi)^{3/2}} \int d^{3} k\,\, \vec{\pi}_{\vec{k}}(\tau) \,\,e^{-i \vec{k}\cdot\vec{x}}, \qquad 
 \vec{A}(\vec{x},\tau) = \frac{1}{(2\pi)^{3/2}} \int d^{3} k \,\, \vec{A}_{\vec{k}}(\tau) \,\,e^{-i \vec{k}\cdot\vec{x}},
 \label{dual9}
 \end{equation}
and it can be inserted into Eq. (\ref{dual8}). The resulting form of the canonical Hamiltonian is:
\begin{equation}
H_{A}(\tau) = \frac{1}{2} \int d^3 k \biggl[ \vec{\pi}_{\vec{k}} \cdot \vec{\pi}_{-\vec{k}} +  \frac{\chi_{E}'}{\chi_{E}}
\biggl( \vec{\pi}_{\vec{k}} \cdot \vec{A}_{-\vec{k}}  +  \vec{\pi}_{-\vec{k}} \cdot \vec{A}_{\vec{k}}\biggr)
+ \frac{k^2}{f} \,\vec{A}_{\vec{k}} \cdot \vec{A}_{-\vec{k}}\biggr].
\label{dual10}
\end{equation}
From Eq. (\ref{dual10}) the corresponding equations or motion are:
\begin{eqnarray}
&& \vec{A}_{\vec{k}}^{\,\prime} = \vec{\pi}_{\vec{k}} + \frac{\chi_{E}'}{\chi_{E}} \vec{A}_{\vec{k}},
\label{dual12}\\
&& \vec{\pi}_{\vec{k}}^{\,\prime} = - \frac{k^2}{f} \,\vec{A}_{\vec{k}} -\frac{\chi_{E}'}{\chi_{E}} \vec{\pi}_{\vec{k}}.
\label{dual13}
\end{eqnarray}
The duality transformation exchanges the canonical fields and the conjugate momenta 
\begin{eqnarray}
&& \chi_{E} \to \frac{1}{\chi_{B}}, \qquad \chi_{B} \to \frac{1}{\chi_{E}}, 
\nonumber\\
&&  \vec{\pi}_{\vec{k}} \to \Pi_{\vec{k}} =- \frac{k}{\sqrt{f}} \, \vec{A}_{\vec{k}}, \qquad \vec{A}_{\vec{k}} \to \vec{{\mathcal A}}_{\vec{k}} = \frac{\sqrt{f}}{k} \vec{\pi}_{\vec{k}},
\label{dual11}
\end{eqnarray}
and it also replaces Eq. (\ref{dual12}) with Eq. (\ref{dual13}) and vice versa. The transformation of Eq. (\ref{dual11}) 
is canonical and the generating functional can be written as:
\begin{equation}
{\mathcal G}[\vec{A},\, \vec{{\mathcal A}},\,\tau] = \int d^{3} k\, \frac{k}{\sqrt{f(\tau)}} \biggl(\vec{{\mathcal A}}_{\vec{k}}\cdot\vec{A}_{\vec{k}}+ \vec{{\mathcal A}}_{-\vec{k}}\cdot\vec{A}_{-\vec{k}}\biggr).
\label{dual11a}
\end{equation}
The transformed Hamiltonian will be given by 
\begin{eqnarray}
H_{A}(\tau) &\to& \overline{H}_{{\mathcal A}} = H_{A} + \frac{\partial {\mathcal G}}{\partial \tau} 
\nonumber\\
&=&  \frac{1}{2} \int d^3 k \biggl[ \vec{\Pi}_{\vec{k}} \cdot \vec{\Pi}_{-\vec{k}} +  \frac{\chi_{E}'}{\chi_{E}}
\biggl( \vec{\Pi}_{\vec{k}} \cdot \vec{{\mathcal A}}_{-\vec{k}}  +  \vec{\Pi}_{-\vec{k}} \cdot \vec{{\mathcal A}}_{\vec{k}}\biggr)
+ \frac{k^2}{f} \,\vec{{\mathcal A}}_{\vec{k}} \cdot \vec{{\mathcal A}}_{-\vec{k}}\biggr],
\label{dual11b}
\end{eqnarray}
where we have used the identity $\chi_{E}'/ \chi_{E} = (\chi_{B}'/\chi_{B} + \sqrt{f}^{\prime}/\sqrt{f})$. 
\subsection{Mode functions and power spectra}
Promoting the canonical fields to quantum operators (i.e. $A_{i} \to \hat{A}_{i}$ and $\pi_{i} \to \hat{\pi}_{i}$) the following
(equal time) commutation relations (in units $\hbar = c =1$) must hold:
\begin{equation}
[\hat{A}_{i}(\vec{x}_{1},\tau),\hat{\pi}_{j}(\vec{x}_{2},\tau)] = i \Delta_{ij}(\vec{x}_{1} - \vec{x}_{2}),\qquad 
\Delta_{ij}(\vec{x}_{1} - \vec{x}_{2}) = \int \frac{d^{3}k}{(2\pi)^3} e^{i \vec{k} \cdot (\vec{x}_{1} - \vec{x}_2)} P_{ij}(k), 
\label{dual17}
\end{equation}
where $P_{ij}(k) = (\delta_{ij} - k_{i} k_{j}/k^2)$. The function $\Delta_{ij}(\vec{x}_{1} - \vec{x}_{2})$ is the transverse generalization of the Dirac delta function
ensuring that both $\vec{E}$ and $\vec{A}$ are divergenceless. The field operators can then be expanded in terms of the corresponding 
mode functions 
\begin{eqnarray}
\hat{A}_{i}(\vec{x},\tau) = \int\frac{d^{3} k}{(2\pi)^{3/2}} \sum_{\alpha} e^{(\alpha)}_{i}(k) \, 
\biggl[ F_{k}(\tau) \, \hat{a}_{k,\alpha} e^{- i \vec{k} \cdot\vec{x}} +  F_{k}^{*}(\tau) \, \hat{a}^{\dagger}_{k,\alpha} e^{ i \vec{k} \cdot\vec{x}}\biggr],
\label{dual18}\\
\hat{\pi}_{i}(\vec{x},\tau) = \int\frac{d^{3} k}{(2\pi)^{3/2}} \sum_{\alpha} e^{(\alpha)}_{i}(k) \, 
\biggl[ G_{k}(\tau) \, \hat{a}_{k,\alpha} e^{- i \vec{k} \cdot\vec{x}} +  G_{k}^{*}(\tau) \, \hat{a}^{\dagger}_{k,\alpha} e^{ i \vec{k} \cdot\vec{x}}\biggr],
\label{dual19}
\end{eqnarray}
where $F_{k}(\tau)$ and $G_{k}(\tau)$ obey:
\begin{eqnarray}
&& F_{k}' = G_{k} + \frac{\chi_{E}'}{\chi_{E}} F_{k},
\label{dual20}\\
&& G_{k}' = -  \frac{k^2}{f} F_{k} - \frac{\chi_{E}'}{\chi_{E}} G_{k}.
\label{dual21}
\end{eqnarray}
Equations (\ref{dual20})--(\ref{dual21}) come from Eqs. (\ref{dual12})--(\ref{dual13}) and the mode functions $F_{k}(\tau)$ and $G_{k}(\tau)$ must also satisfy the Wronskian normalization condition which follows from the canonical commutators together with the expansions (\ref{dual18}) and (\ref{dual19}):
\begin{equation}
F_{k}(\tau)\,G_{k}^{*}(\tau) - F_{k}^{*}(\tau)\,G_{k}(\tau) =i.
\label{dual22}
\end{equation}
The equations for the mode functions can be decoupled with the usual manipulations: 
\begin{eqnarray}
&& F_{k}'' + \biggl[ \frac{k^2}{f}   - \frac{\chi_{E}''}{\chi_{E}} \biggr]F_{k} =0,
\label{mode1aa}\\
&& \overline{G}_{k}''   + \biggl[ \frac{k^2}{f} - \biggl(\frac{1}{\chi_{B}}\biggr)^{\prime\prime} \chi_{B}\biggr] \overline{G}_{k} =0,
\label{mode2aa}
\end{eqnarray}
where  $\overline{G}_{k} = \sqrt{f} G_{k}$. In terms of $F_{k}$ and $G_{k}$ the magnetic and the electric power spectra are\footnote{The factor $1/f$ in Eq. (\ref{PB}) may appear at first sight odd but it comes from the 
correct relation between the magnetic field and the canonical normal mode $\vec{A}$.}
\begin{eqnarray}
P_{B}(k,\tau) &=& \frac{k^{5}}{2\, \pi^2\, a^4(\tau) \, f(\tau)} \, |F_{k}(\tau)|^2, 
\label{PB}\\
P_{E}(k,\tau) &=& \frac{k^3}{2\, \pi^2\, a^4(\tau)} \,  |G_{k}(\tau)|^2.
\label{PE}
\end{eqnarray}
The correlators of the rescaled fields in Fourier space are given by:
\begin{eqnarray}
&& \langle B_{i}(\vec{k},\tau)\, B_{j}(\vec{p},\tau) \rangle = \frac{2\pi^2}{k^3}\, P_{B}(k,\tau)\, P_{ij}(k)  \,\delta^{(3)}(\vec{k} + \vec{p}),
\label{cc1}\\
&& \langle E_{i}(\vec{k},\tau)\, E_{j}(\vec{p},\tau) \rangle = \frac{2\pi^2}{k^3}\, P_{E}(k,\tau)\, P_{ij}(k)  \,\delta^{(3)}(\vec{k} + \vec{p}),
\label{cc2}
\end{eqnarray}
where $P_{ij}(k)$ has been defined after Eq. (\ref{dual17}) while $B_{i}(\vec{k},\tau)$ and $E_{i}(\vec{k},\tau)$ are, strictly 
speaking, field operators in Fourier space but can be also viewed as classical stochastic variables.
From Eqs. (\ref{cc1})--(\ref{cc2}) and (\ref{enm3})  the properly normalized energy density is 
\begin{equation}
\rho_{E} + \rho_{B} = \int \frac{d k}{k} \biggl[P_{B}(k,\tau) + P_{E}(k,\tau) \biggr],
\label{enm4}
\end{equation}
where the $4\pi$ factor disappeared because it has been included in the canonical redefinition of the fields. 
\subsection{Power spectra and duality}
Under the duality transformation $\chi_{E} \to 1/\chi_{B}$ and $\chi_{B} \to 1/\chi_{E}$, Eqs. (\ref{dual20}) and Eq. (\ref{dual21}) are exchanged provided
\begin{equation}
 G_{k}\to - \frac{k}{\sqrt{f}}\, F_{k},\qquad F_{k} \to \frac{\sqrt{f}}{k} G_{k} ,
\label{DD2}
\end{equation}
This property is a consequence of Eq. (\ref{dual11}) but it can be directly verified.
Indeed,  using Eq. (\ref{DD2}),  Eq. (\ref{dual20}) transforms as:
\begin{equation}
F_{k}' = G_{k} + \frac{\chi_{E}'}{\chi_{E}} F_{k}\to \biggl(\frac{\sqrt{f}}{k}\,G_{k}\biggr)^{\prime} = - \frac{k}{\sqrt{f}} F_{k} + \biggl(\frac{1}{\chi_{B}}\biggr)^{\,\prime} \, \chi_{B}\, \biggl(\frac{\sqrt{f}}{k}G_{k}\biggr).
\label{expltr}
\end{equation}
After performing the time derivative on the left-hand side of Eq. (\ref{expltr}), both sides of the equation can be multiplied by $k/\sqrt{f}$; Eq. (\ref{expltr}) becomes:
\begin{equation}
G_{k}' = - \frac{k^2}{f}  F_{k} - \biggl(\frac{\chi_{B}'}{\chi_{B}} + \frac{\sqrt{f}\,'}{\sqrt{f}}\biggr) G_{k},
\label{expltr3}
\end{equation}
which coincides exactly with Eq. (\ref{dual21}) if we recall that, by definition of $f$,  $\chi_{\mathrm{B}} \sqrt{f} \equiv \chi_{E}$.
Using the transformations of Eq. (\ref{DD2}), the spectra of Eqs. (\ref{PB}) and (\ref{PE}) are interchanged, i.e. 
\begin{equation}
P_{B}(k,\tau) \to P_{E}(k,\tau),\qquad P_{E}(k,\tau) \to P_{B}(k,\tau).
\label{expltr4}
\end{equation}
Equation (\ref{expltr4}) relates different dynamical regimes in the evolution 
of $\chi_{E}$ and $\chi_{B}$.  In summary, since 
Eqs. (\ref{mode1aa}) and (\ref{mode2aa})
are invariant under the generalized duality transformation, also the evolution equations of the 
mode functions are exchanged by duality. This conclusion implies that the magnetic and electric power 
spectra are exchanged by the action of the duality symmetry in such a way that the total energy density 
is left unaltered.
\renewcommand{\theequation}{4.\arabic{equation}}
\setcounter{equation}{0}
\section{Inflationary magnetic and electric power spectra}
\label{sec4}
\subsection{General considerations}
For an explicit solution of Eqs. (\ref{mode1aa})--(\ref{mode2aa}) the susceptibilities shall be parametrized as\footnote{Although the variable $y$ can be explicitly expressed either in terms of the conformal time coordinate or in terms of the total number of efolds elapsed since $\tau_{i}$ (i.e. $\ln{y(\tau)} = - N_{t}$), the latter parametrization appears to be more useful than the former when dealing with phenomenological considerations as we shall point out in sec. \ref{sec5}.}
\begin{equation}
\chi_{E}(y) = y^{1/2 - \nu},\qquad  \chi_{B}(y) = y^{1/2 - \nu + \mu}, \qquad f(y) = y^{- 2\mu},
\label{pumps}
\end{equation}
where $y(\tau) = (- \tau/\tau_{i})$ and $\tau_{i}$ marks the initial time of the evolution of the various pump fields and the 
relevant dynamical evolution occurs for $\tau > - \tau_{i}$. The parametrization given in Eq. (\ref{pumps}) is monotonic even if this assumption can be easily relaxed within the same scheme\footnote{
In the case of bouncing models of magnetogenesis the evolution may also be non-monotonic, as argued in the past \cite{notula}.}. During the quasi-de Sitter stage of expansion the following standard 
relations hold between the expansion rates in conformal (i.e. ${\mathcal H} = a^{\prime}/a$) and in 
cosmic time (i.e. $H =\dot{a}/a$):
\begin{equation}
{\mathcal H} = a H = - \frac{1}{(1 - \epsilon) \tau}, \qquad \epsilon = - \frac{\dot{H}}{H^2},
\label{HH}
\end{equation}
where the overdot denotes a derivation with respect to the cosmic time coordinate and $\epsilon$ is 
standard slow-roll parameter.  Defining $\alpha = 1/2 -\nu$ and focussing on the case $\mu >0$ there are three distinct regions in the $(\alpha,\, \mu)$ plane. If $\alpha > 0$,  $\chi_{E}$ and $\chi_{B}$  are both decreasing. Conversely, in the region $(\alpha + \mu) > 0$ and $\alpha <0$ (i.e.  $- \mu < \alpha <0$), $\chi_{E}$ increases while $\chi_{B}$ decreases. Finally $\chi_{E}$ and $\chi_{B}$ are both increasing as a function of $\tau$ in the region $\alpha < - \mu < 0$.  If  $\mu < 0$ (or if the sign of $\mu$ is flipped in Eq. (\ref{pumps})) the $(\alpha,\, \mu)$ plane is still divided in three regions. More specifically  $\chi_{E}$ and $\chi_{B}$  are both decreasing for $0< \alpha  <\mu $. Conversely, in the region $0 < \mu  <\alpha $, $\chi_{E}$ decreases while $\chi_{B}$ increases; finally, $\chi_{E}$ and $\chi_{B}$ are both increasing in the region $\alpha  < 0$. If we relate $1/\chi_{C}$ (with $C= B,\, E$) to the gauge coupling, the increase  of $\chi_{C}$ implies a decrease of the gauge coupling and vice versa.

 Equation (\ref{HH}) holds in the case of 
conventional inflationary models (see e.g. \cite{rev6}) where the Universe evolves from strong gravitational coupling to weak gravitational coupling, i.e. the space-time curvature is maximal at the onset of inflation 
and gets smaller during reheating. It is fair to say that the potential drawbacks of magnetogenesis coincide with the potential drawbacks of conventional models of inflation which are, typically, not geodesically complete in their past history. 
The considerations reported here can be easily extended to the case of bouncing models (see e.g. \cite{rev5} for this terminology) evolving from weak gravitational coupling to strong gravitational coupling, i.e. the space-time curvature is small initially and gets larger at the reheating.  

The parametrization of Eq. (\ref{pumps}) is general enough to encompass all the physically interesting cases and the aim of the forthcoming considerations is to relate the electric and magnetic power spectra to the evolution of the susceptibilities.
In other words, given a sufficiently general parametrization for the evolution of the susceptibilities such as 
the one of Eq. (\ref{pumps}) which are the corresponding power spectra obtainable during a phase of quasi-de Sitter evolution? Are they phenomenologically relevant? These are some of the questions addressed in the present and in the following section. 

\subsection{Analytic solutions for the mode functions}

Inserting Eq. (\ref{pumps}) into Eqs. (\ref{mode1aa}) and (\ref{mode2aa}) and defining $z = (-\tau)$, the resulting pair of equations is:
\begin{eqnarray}
&& F_{k}'' + \frac{1 - 2 p_{F}}{z} F_{k}' + \biggl[ \gamma_{F}^2 \, q^2 \, z^{ 2 q - 2} + \frac{p_{F}^2 - \sigma^2 \, q^2}{z^2} \biggr] F_{k}=0, 
\label{fex}\\
&& \overline{G}_{k}'' + \frac{1 - 2 p_{G}}{z} \overline{G}_{k}' + \biggl[ \gamma_{G}^2 \, q^2 \, z^{ 2 q - 2} + \frac{p_{G}^2 - \rho^2 \, q^2}{z^2} \biggr] \overline{G}_{k}=0,
\label{gex}
\end{eqnarray}
where $\overline{G}_{k}(\tau) = \sqrt{f(\tau)} \, G_{k}(\tau)$, and 
\begin{equation}
p_{F} = 1/2, \qquad  p_{G} = 1/2,\qquad q = (1 + \mu), \qquad \gamma_{F}= \gamma_{G} = \frac{k}{\tau_{i}^{\mu} \, |1 + \mu|}.
\label{def}
\end{equation}
Equations (\ref{fex}) and (\ref{gex}) are different from the analog equations obtainable in the case 
when the susceptibilities are coincident. The solution of Eqs. (\ref{fex})--(\ref{gex}) can be obtained in terms of two linear combinations of Bessel functions \cite{abr,tric} with indices $\sigma$ and $\rho$ denoted hereunder  by ${\mathcal C}_{\sigma}$ and ${\mathcal C}_{\rho}$:
\begin{eqnarray} 
&& F_{k}(z) = z^{p_{F}} {\mathcal C}_{\sigma}(\gamma_{F} \, z^{q}), \qquad \qquad \sigma=\frac{\nu}{1+ \mu},
\label{fsol}\\
&& \overline{G}_{k}(z) = z^{p_{G}} {\mathcal C}_{\rho}(\gamma_{G} \, z^{q}), \qquad \qquad \rho= \sigma -1.
\label{gsol}
\end{eqnarray}
According to Eq. (\ref{dual21}), the relation between $G_{k}$ and $F_{k}'$ is given by $G_{k} = F_{k}' - (\chi_{E}'/\chi_{E}) F_{k}$.  Imposing the quantum mechanical normalization, 
Eqs. (\ref{fsol})-(\ref{gsol}) are expressible in terms of Hankel functions of first 
kind \cite{abr,tric}:
\begin{eqnarray}
&& F_{k}(\tau) =\frac{ {\mathcal D}}{ \sqrt{2 k/\sqrt{f(\tau)}}}\sqrt{- k \tau/\sqrt{f(\tau)}}\,\, H_{\sigma}^{(1)}\biggl( - \frac{k\, \tau/\sqrt{f(\tau)}}{| 1 + \mu| } \biggr),
\label{fsol1}\\
&& G_{k}(\tau) = -  {\mathcal D} \sqrt{\frac{k/\sqrt{f(\tau)} }{2\, }} \, \sqrt{ - k \tau/\sqrt{f(\tau)}}\,  H_{\rho}^{(1)}\biggl( - \frac{k\,\tau/\sqrt{f(\tau)}}{| 1 + \mu|}\biggr),
\label{fsol2}
\end{eqnarray}
where $|{\mathcal D}|^2 = \pi/(2 | 1 + \mu|)$; $\sigma$ and $\rho= \sigma-1$ have been already defined in Eqs. (\ref{fsol})--(\ref{gsol}). Equations (\ref{fsol1}) and (\ref{fsol2}) satisfy the Wronskian normalization condition of Eq. (\ref{dual22}). The absolute values $|1+\mu|$ guarantee that the results are still valid when $\mu \to - \mu$. 

\subsection{Explicit form of the power spectra}

Inserting Eqs. (\ref{fsol1})--(\ref{fsol2}) into  Eqs. (\ref{PB})--(\ref{PE}) the magnetic and the electric power spectra become:
\begin{eqnarray}
&& P_{B}(k,\tau,\sigma,\mu) = \frac{H^4}{8 \pi \, |1+ \mu|} \frac{(- k \tau)^5}{f^2(\tau)} \, \biggl|H_{\sigma}^{(1)} \biggl(\frac{- k\tau/\sqrt{f(\tau)} }{|1 + \mu|}\biggr) \biggr|^2,
\label{sp1}\\
&& P_{E}(k,\tau,\sigma,\mu) = \frac{H^4}{8 \pi \, |1+\mu|} \frac{(- k \tau)^5}{f^2(\tau)} \, \biggl|H_{\sigma-1}^{(1)} \biggl(\frac{- k\tau/\sqrt{f(\tau)}}{|1 + \mu|} \biggr) \biggr|^2. 
\label{sp2}
\end{eqnarray}
Equations (\ref{sp1}) and (\ref{sp2}) are exchanged\footnote{For a Hankel function with generic index $\alpha$ and argument $z$ we have that $|H^{1}_{\alpha}(z)|^2 = |H^{1}_{-\alpha}(z)|^2$. Thanks to this property
it is possible to show that the electric and magnetic power spectra  are exchanged when $\sigma \to \tilde{\sigma} = 1 - \sigma$. This invariance is related to the duality symmetry.} if  $\sigma \to \tilde{\sigma} = 1 - \sigma$. 
In terms of the two dimensionless variables $x = - k \tau$ and $y = (-\tau/\tau_{i})$, Eqs. (\ref{sp1}) and (\ref{sp2}) are
\begin{eqnarray}
 && P_{B}(x,\, y,\, \sigma,\,\mu) = \frac{H^4}{8 \pi |1 + \mu|} \, \, x^{5}\,y^{2 \mu}\,  \biggl|H^{(1)}_{\sigma}\biggl( \frac{x\, y^{\mu}}{| 1 + \mu|}\biggr)\biggr|^2,
 \label{sp1a}\\
 && P_{E}(x,\, y,\,\sigma,\,\mu) =\frac{H^4}{8 \pi |1 + \mu|} \, \, x^{5}\, y^{2 \mu}  \biggl|H^{(1)}_{\sigma-1}\biggl( \frac{x\, y^{\mu}}{| 1 +\mu|}\biggr)\biggr|^2.
 \label{sp2a}
 \end{eqnarray}
When the relevant wavelengths are larger than the Hubble 
radius it is practical to introduce yet another variable defined as $w = x \, y^{\mu}$. In the $(y,\, w)$ plane,  Eqs. (\ref{sp1a}) and (\ref{sp2a}) read:
 \begin{eqnarray}
 && P_{B}(y,\,w,\,\sigma,\,\mu) = \frac{H^4}{8 \pi\, |1+\mu|} \, w^5\, y^{-3 \mu}\,   \biggl|H^{(1)}_{\sigma}\biggl( \frac{w}{| 1 + \mu|\, }\biggr)\biggr|^2,
\label{sp1b}\\
 && P_{E}( y,\,w,\,\sigma,\,\mu) = \frac{H^4}{8 \pi\, |1+\mu|} \, w^5\, y^{-3 \mu}\,   \biggl|H^{(1)}_{\sigma-1}\biggl( \frac{w}{| 1 + \mu|\, }\biggr)\biggr|^2.
 \label{sp2b}
 \end{eqnarray}
Wavelengths larger than the Hubble radius correspond to the condition $|k /\sqrt{f}|< {\mathcal H}$.  The case $\mu=-1$ is singular since,  in this case, $k^2/f$ and $\chi_{E}''/\chi_{E}$ evolve roughly at the same rate. This implies that the modes 
that are larger than the Hubble rate at $\tau_{i}$ will never reenter while the modes inside the Hubble radius at $\tau_{i}$ will never exit.
In the limit  $w \ll 1$ the corresponding wavelengths are larger than the Hubble radius and the power spectra of Eqs. (\ref{sp1b})--(\ref{sp2b}) become
 \begin{eqnarray}
P_{B}(x,\,y,\, \sigma) &=& H^4 \,\,{\mathcal Q}_{B}(\sigma,\mu) \,  \,x^{5 - 2 |\sigma|} \, \, y^{ - 2 \mu ( |\sigma|-1)},
\label{sp1c}\\
P_{E}(x,\,y,\, \sigma) &=& H^4 \,\,{\mathcal Q}_{E}(\sigma,\mu) \,  \,x^{5 - 2 |\sigma\, -\,1 |} \, \, y^{ - 2 \mu ( |\sigma-1|-1)},
\label{sp2c}
\end{eqnarray}
where 
\begin{eqnarray}
{\mathcal Q}_{B}(\sigma,\,\mu) &=& \frac{\Gamma^2(|\sigma|)}{\pi^3} \, 2^{ 2 |\sigma| -3} \, |1+ \mu|^{ 2 |\sigma|-1},
\nonumber\\
{\mathcal Q}_{E}(\sigma,\,\mu) &=&  \frac{\Gamma^2(|\sigma-1|)}{\pi^3} \, 2^{ 2 |\sigma-1| -3} \, |1+ \mu|^{ 2 |\sigma-1|-1}.
\label{QQ}
\end{eqnarray}
The amplitude of the spectra of Eqs. (\ref{sp1c})--(\ref{sp2c}) depends on $\sigma$ and on $\mu$: on the one hand $\sigma$ is defined in terms of $\mu$ and $\nu$ (i.e.  $\sigma = \nu/(1+\mu)$), on the other hand $\mu$ controls the overall suppression or enhancement of the spectrum through the $y$-dependent prefactor that is related to the total number of efolds.  To proceed further a more transparent parametrization of the spectral indices is desirable. 

\subsection{Spectral indices}
The magnetic and the electric spectral indices are defined as:
\begin{equation}
n_{B} - 1 = \frac{\partial P_{B}(x,\,y,\,\sigma,\, \mu)}{\partial \ln{x}}, \qquad n_{E} - 1 = \frac{\partial P_{E}(x,\,y,\,\sigma,\,\mu)}{\partial \ln{x}},
\label{slope}
\end{equation}
where the scale-invariant limits correspond to $n_{E} \to 1$ and $n_{B} \to 1$. The power 
spectrum of curvature perturbations ${\mathcal P}_{{\mathcal R}}(k)$ is assigned (see e.g. \cite{wmap1,wmap2,wmap3})
within the same conventions 
\begin{equation}
{\mathcal P}_{{\mathcal R}}(k) = {\mathcal A}_{{\mathcal R}} \biggl(\frac{k}{k_{\mathrm{p}}}\biggr)^{n_{\mathrm{s}}-1}, \qquad k_{\mathrm{p}} = 0.002\,\, \mathrm{Mpc}^{-1}, 
\label{PSR}
\end{equation}
where ${\mathcal A}_{{\mathcal R}}$ (the spectral amplitude at the pivot scale $k_{\mathrm{p}}$) determines the inflationary rate 
of expansion and enters directly the amplitude of the magnetic and electric power spectra (see sec. \ref{sec5}); $n_{\mathrm{s}}$ is the scalar 
spectral index.
As implied by the absolute values appearing in Eqs. (\ref{sp1c})--(\ref{sp2c}), the power spectra have three different analytic forms depending on the values of $\sigma$: 
\begin{itemize}
\item{} if  $\sigma > 1$ the magnetic and 
the electric spectral indices  are, respectively $n_{B} = 6 - 2 \sigma$ and $n_{E} = 8 - 2\sigma$; the consistency between the two indices implies, in this 
region, $n_{E} = n_{B}+2$;
\item{} if $0 < \sigma < 1$ the slope of the electric power spectrum is unchanged in 
comparison with the previous case; on the contrary $n_{E}$ is given by  $n_{E} = 4 + 2 \sigma$; the consistency between $n_{E}$ and $n_{B}$ implies, in this case, $n_{E} = 10 - n_{B}$;
\item{} if $\sigma < 0$ the magnetic and the electric spectral indices are, respectively, $n_{B} = 6 + 2\sigma$ and $n_{E} = 4 + 2\sigma$, implying $n_{E} = n_{B} - 2$.
\end{itemize} 
Consider now the limit $\mu \to 0$  (see also Eq. (\ref{limit1})): when $\mu \to 0$,
$f \to 1$,  $\chi_{E} = \chi_{B} = \sqrt{\lambda}$ and $\sigma \to \nu$. All the relations between the spectral indices and $\sigma$ deduced in the previous list remain true in the limit $\mu \to 0$ provided $\sigma$ is replaced by $\nu$. When $\mu \neq 0$ the spectral indices and the corresponding amplitudes are determined by not only by $\nu$ but also by $\mu$: what was a line in the parameter space connecting $n_{B}$ (or $n_{E}$) to $\nu$ becomes now a plane.   This 
is, in a nutshell, the rationale for the widening of the parameter space of the model. 
\subsection{Regions in the parameter space}
Although the parameter 
space of the model can be charted either in the $(\mu,\,\nu)$ plane or in the $(\mu,\, \sigma)$ plane, the latter parametrization turns out to be 
more useful than the former since the spectral indices have a simpler dependence in terms of $\sigma$. Moreover since $\nu = \sigma ( 1 + \mu)$, $\nu$ can be eliminated 
from the rate of variation of the susceptibilities of Eq. (\ref{pumps}) so that $\ln{\chi_{E}} = [1/2 - \sigma (1 + \mu)] \ln {y}$ and  $\ln{\chi_{B}} = [1/2 + \mu -  \sigma (1 + \mu)] \ln {y}$. From these expressions we can say that $\chi_{E}$ and $\chi_{B}$ are both decreasing during the quasi-de Sitter stage of expansion provided $[1/2 - \sigma (1 + \mu)] >0$ and $[1/2 + \mu -  \sigma (1 + \mu)]>0$. With similar logic the entire parameter space can be discussed. 
\begin{figure}[!ht]
\centering
\includegraphics[height=7cm]{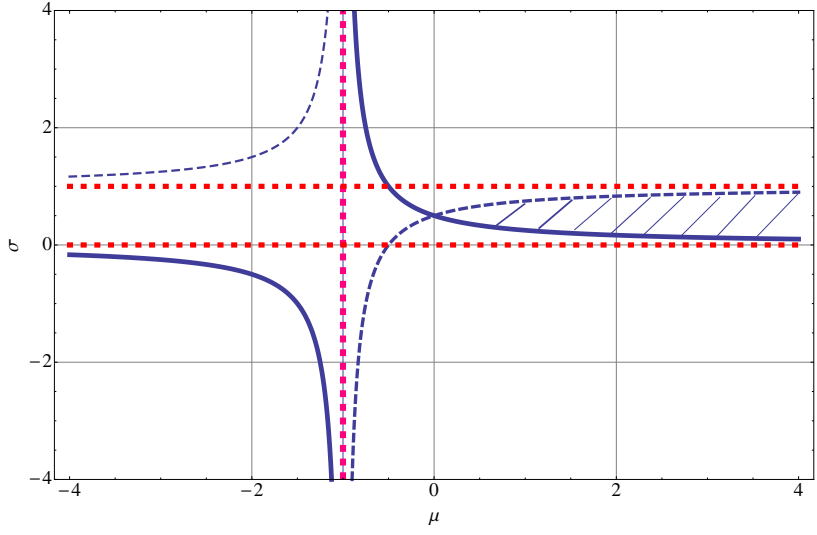}
\caption[a]{The regions of the $(\sigma,\, \mu)$ plane are illustrated.}
\label{Figure0}      
\end{figure}
In Fig. \ref{Figure0} the various regions of the $(\sigma,\, \mu)$ plane are reported. 
Below the two {\em dashed} branches of hyperbola 
$\chi_{B}$ is decreasing. Similarly, below the two {\em full} branches of hyperbola
 $\chi_{E}$ is decreasing. Above the same curves (either dashed or full) 
the situation is reversed and the corresponding susceptibilities increase rather than decreasing. 

The shaded area of Fig. \ref{Figure0} (bounded from above by the dashed
hyperbola and from below by the full hyperbola) describes an intermediate situation: in this region $\chi_{B}$ decreases while $\chi_{E}$ increases. 
In Fig. \ref{Figure0} the two horizontal dotted lines are the asymptotes of the two 
hyperbolae (i.e. $\sigma=0$ and $\sigma =1$)  but they 
are also the boundaries of the three regions characterizing the different values of the spectral indices discussed in the 
list of items of the previous subsection.
The line $\mu=-1$ (i.e. the common vertical asymptote of both hyperbolae)  
has been already discussed after Eqs. (\ref{sp1b})--(\ref{sp2b}):
when $\mu = -1$ the pumping action due to the susceptibility and to the refractive index are 
exactly balanced (i.e. $1/f \simeq \chi_{E}''/\chi_{E}$) and  both proportional to $\tau^{-2}$. 

\renewcommand{\theequation}{5.\arabic{equation}}
\setcounter{equation}{0}
\section{Phenomenology}
\label{sec5}
\subsection{Power spectra in critical units}
From Eq. (\ref{enm4}) the electric and magnetic energy densities in critical units are:
\begin{equation}
\Omega_{B}(x,\, y,\,\sigma,\, \mu)  = \frac{8 \pi}{3} \, \frac{P_{B}(x,\, y,\, \sigma,\, \mu)}{  H^2 \, M_{\mathrm{P}}^2}, \qquad \Omega_{E}(x,\, y,\, \sigma,\, \mu)   = \frac{8 \pi}{3} \frac{P_{E}(x,\, y,\, \sigma,\, \mu)}{ H^2 \,M_{\mathrm{P}}^2},
\label{om1}
\end{equation}
where $M_{P}$ is the Planck mass. Inserting Eqs. (\ref{sp1c}) and (\ref{sp2c}) into Eq. (\ref{om1}) and recalling the notations 
of Eqs. (\ref{HH}) and (\ref{PSR}), Eq. (\ref{om1}) leads to the following pair of equations: 
\begin{eqnarray}
\Omega_{B}(x,\, y,\, \sigma,\,\mu)  &=& \frac{8 \pi^2}{3} \,\, \epsilon \,\, {\mathcal A}_{{\mathcal R}} \, {\mathcal Q}_{B}(\sigma,\,\mu)
 \,\, x^{5 - 2 |\sigma|} \,\, y^{- 2 \mu( |\sigma | -1)},
\label{om2}\\
\Omega_{E}(x,\, y,\, \sigma,\, \mu)  &=& \frac{8 \pi^2}{3} \, \epsilon \, {\mathcal A}_{{\mathcal R}} \, {\mathcal Q}_{E}(\sigma,\,\mu) 
\,\, x^{5 - 2 |\sigma-1|} \,\, y^{-2 \mu( |\sigma -1 | -1)}.
\label{om3}
\end{eqnarray}
In Eqs. (\ref{om2})--(\ref{om3}) the inflationary Hubble rate has been expressed in terms of the amplitude of adiabatic curvature perturbations.
The fiducial set of cosmological parameters\footnote{Using the standard terminology $\Omega_{\mathrm{b}0}$, $\Omega_{\mathrm{c}0}$ and  $\Omega_{\mathrm{de}0}$ 
are the critical fractions of baryons, dark matter and dark energy; $h_{0}$ is the Hubble rate at the present time and in units of $100\, \mathrm{km/sec\, Mpc}$;
$n_{\mathrm{s}}$ and $\epsilon_{\mathrm{re}}$ are, respectively, the spectral index of curvature perturbations and the optical depth at 
reionization.} used hereunder comes from the comparison of the concordance paradigm with the WMAP 9 yr data alone \cite{wmap1} (see also \cite{wmap2,wmap3}):
\begin{equation}
( \Omega_{\mathrm{b}0}, \, \Omega_{\mathrm{c}0}, \Omega_{\mathrm{de}0},\, h_{0},\,n_{\mathrm{s}},\, \epsilon_{\mathrm{re}}) \equiv 
(0.0463,\, 0.233,\, 0.721,\,0.700,\, 0.972,\,0.089),
\label{ppp1}
\end{equation}
with ${\mathcal A}_{{\mathcal R}} = 2.41\times 10^{-9}$. The combinations of other data sets lead to slight differences in the pivotal parameters but these 
differences have no relevance in the present context. For instance, using the data  of the baryon acoustic oscillations (see, e.g. \cite{SDSS}) in combination with the WMAP 9 yr data, the six parameters of Eq. (\ref{ppp1}) are modified at the level of the few percent  and ${\mathcal A}_{{\mathcal R}} = 2.35\times 10^{-9}$.
Another set of concordance parameters is obtained by combining the WMAP 9 yr data with 
the direct determinations of the Hubble rate giving  ${\mathcal A}_{{\mathcal R}} = 2.45\times 10^{-9}$. The differences in the values of ${\mathcal A}_{{\mathcal R}}$ are immaterial for the present considerations. The same comment holds for the values of $\epsilon$ whose upper limits range 
from $\epsilon < 0.023$ in the case of WMAP 9 yr data alone to $\epsilon< 0.0081$ when the WMAP 9 yr data are combined with all the 
other data (see e.g. \cite{SDSS,SN1,SN2,perci,act,spt,snls3}). The Planck explorer data, at least in their current release, do not lead to crucial differences 
in the determinations of the concordance parameters and cannot be used alone but must  be combined, in some way, with the WMAP data.

\subsection{Dependence on the number of efolds}
The variable $x$ appearing in Eqs. (\ref{om1}), (\ref{om2}) and (\ref{om3}) can be expressed as: 
\begin{eqnarray}
x &=& \frac{k}{(1 -\epsilon) a H} = \frac{k}{H_{0} } e^{ - N_{\mathrm{max}}} \biggl[1 + \epsilon + {\mathcal O}(\epsilon^2)\biggr]
\label{xx}\\
&\simeq& 6.35\times 10^{-24} \, \biggl(\frac{k}{\mathrm{Mpc}^{-1}}\biggr) \, \biggl(\frac{h_{0}}{0.7} \biggr)^{-1}\,\,\biggl(\frac{\epsilon}{0.01}\biggr)^{-1/4} \, 
\biggl(\frac{{\mathcal A}_{\mathcal R}}{2.41\times 10^{-9}} \biggr)^{-1/4},
\label{ex}
\end{eqnarray}
where $H_{0} = 100 \,h_{0} \,\mathrm{Mpc}^{-1}\, \mathrm{km}/\mathrm{sec}$ is the present value of the Hubble rate and 
$N_{\mathrm{max}}$ is the maximal number of efolds which are today accessible to our observations \cite{efolds}. In practice $N_{\mathrm{max}}$ is determined by fitting the redshifted inflationary event horizon inside the present Hubble radius $H_{0}^{-1}$:
\begin{equation}
e^{N_{\mathrm{max}}} = ( 2 \, \pi  \, \epsilon \, {\mathcal A}_{{\mathcal R}}\, \Omega_{\mathrm{R}0})^{1/4} \,\, \biggl(\frac{M_{\mathrm{P}}}{ H_{0} }\biggr)^{1/2} \biggl(\frac{H_{r}}{H}\biggr)^{\gamma -1/2},
\label{NN1}
\end{equation}
where $\Omega_{\mathrm{R}0}$ is the present critical fraction of radiation (in the concordance model $h_{0}^2 \Omega_{\mathrm{R}0} 
= 4.15 \times 10^{-5}$). From Eq. (\ref{ppp1}) and in the 
sudden reheating approximation we have  $N_{\mathrm{max}} \simeq 63.25 + 0.25 \ln{\epsilon}$ which is 
numerically close to the minimal number of efolds $N_{\mathrm{min}}$ needed to solve the kinematic 
problems of the standard cosmological model (i.e. $N_{\mathrm{min}} \simeq N_{\mathrm{max}}$). 

Because of the possibility of a delayed reheating the value of $N_{\mathrm{max}}$ suffers of a certain degree of theoretical uncertainty which can be roughly quantified in $15$ efolds.
Indeed,  Eq. (\ref{NN1}) assumes that the reheating is concluded at a typical scale $H_{r} \geq 10^{-44}\, M_{\mathrm{P}}$ i.e.  just prior to the formation of the light nuclei.  The expansion rate during the intermediate phase between $H$ and $H_{r}$ 
is controlled by $\gamma$ which can be either smaller than $1/2$ or larger than $1/2$; the case  $\gamma =1/2$ corresponds to the sudden
reheating approximation when the intermediate phase is absent from Eq. (\ref{NN1}).
If $\gamma - 1/2 >0$ (as it happens if $\gamma = 2/3$ when the post-inflationary background is dominated by dust) $N_{\mathrm{max}}$ diminishes in comparison with the case when $H=H_{r}$. Conversely, if $\gamma - 1/2 <0$ (as it happens if $\gamma = 1/3$ when the post-inflationary background is dominated by stiff sources), $N_{\mathrm{max}}$ increases. The maximal increase (of about $15$ efolds) occurs when the post-inflationary evolution is 
dominated by stiff sources down to the epoch of formation of light nuclei.
Moreover, defining $N_{t}$ as the total number of efolds elapsed since $\tau_{i}$, if  $N_{\mathrm{t}} > N_{\mathrm{max}}$, the redshifted value of the inflationary event horizon is larger than the present value of the Hubble radius. 

To summarize the previous considerations, the pivot values considered in the numerical examples will be $N_{\mathrm{max}}= 63.25 + 0.25 \ln{\epsilon}$ and $N_{t} = 80$. These values are both conservative and illustrative given the unavoidable uncertainty about the total duration of the inflationary phase and, to some extent, on the post-inflationary expansion rate. 

\subsection{Post-inflationary evolution}

For the standard thermal history with sudden reheating, the conductivity $\sigma_{\mathrm{c}}$ jumps at a finite value at the end of inflation and the continuity of the electric and magnetic fields implies that the amplitude of the electric power spectrum gets 
suppressed, at a fixed time, as $(k/\sigma_{\mathrm{c}})^2$ in comparison with its magnetic counterpart \cite{DT9}. Both power spectra are exponentially suppressed, for sufficiently large $k$, as $\exp{[-  2(k^2/k_{\sigma}^2)]}$ 
 where $k_{\sigma}^{-2}  = \int_{\tau_{\sigma}}^{\tau} \, d\tau' /[4\pi \sigma_{\mathrm{c}}(\tau')]$. The evaluation 
of $k_{\sigma}$ is complicated by the fact that the integral extends well after $\tau_{\sigma}$. This estimate
can be made rather accurate by computing the transport coefficients of the plasma in different regimes \cite{cond}.  By taking $\tau = \tau_{\mathrm{eq}}$ 
the following approximate expression holds:
\begin{equation}
\biggl(\frac{k}{k_{\sigma}}\biggr)^2 \simeq \frac{10^{-26}}{ \sqrt{2 \, h_{0}^2 \Omega_{\mathrm{M}0} (z_{\mathrm{eq}}+1)}} \, \biggl(\frac{k}{\mathrm{Mpc}^{-1}} \biggr)^2,
\label{CC4}
\end{equation}
where $\Omega_{\mathrm{M}0}= \Omega_{\mathrm{c}0} + \Omega_{\mathrm{b}0}$ and $z_{\mathrm{eq}} \simeq 3200$. Eq. (\ref{CC4}) shows that $\exp{[ - 2(k/k_{\sigma})^2]}$ is so close to $1$ to give negligible suppression for ${\mathcal O}(10^{-4}\, \mathrm{Mpc}^{-1}) \leq k \leq {\mathcal O}(\mathrm{Mpc}^{-1})$ where the 
magnetogenesis considerations apply.  The effect of the conductivity 
is particularly important for blue (i.e. $n_{B} \geq 1$) or violet (i.e. $n_{B} \gg 1$) power spectra since, in these cases, the back-reaction bounds are more constraining at small scales (i.e. large $k$-modes). Equation (\ref{CC4}) would imply that $k_{\sigma} \simeq 10^{13} \, \, \mathrm{Mpc}^{-1}$ but, to be on the safe side,  we shall be even more demanding and require, in the case of increasing power spectra, the back-reaction constraints are met at en even smaller length-scale which is the one corresponding to $x \simeq 1$, i.e. $k \sim a \, H$.

In summary we can say that there are two different physical situations:
\begin{itemize}
\item{} the case of blue or violet spectra (i.e. $n_{B} >1$): in this case the most relevant constraint come from the scales affected by 
the conductivity; to be conservative the constraints shall be applied for $k\sim a H$ even if over these scales the power spectra are 
exponentially suppressed;
\item{} the case of red spectra (i.e. $n_{B} <1$) in this case the most relevant constraints come from large wavelengths or, in equivalent terms, from small wavenumbers in the range $ 10^{-4}\, \mathrm{Mpc}^{-1}\leq k \leq \mathrm{Mpc}^{-1}$.
\end{itemize}
\subsection{The case $\sigma >1$} 
Inserting Eqs. (\ref{sp1c})--(\ref{sp2c}) into  Eqs. (\ref{om2})--(\ref{om3}),  the explicit form of
the power spectra for $\sigma >1$ is:
\begin{eqnarray} 
\Omega_{B}(k,\, N_{t},\, \sigma,\, \mu) &=& \frac{8 \pi^2}{3} \, {\mathcal A}_{{\mathcal R}} \, \, \epsilon \,\, {\mathcal Q}_{B}(\sigma,\,\mu) \biggl(\frac{k}{a H}\biggr)^{5 - 2 \sigma} \,\, e^{2 \mu N_{t} ( \sigma  -1)},
\label{om2a}\\
\Omega_{E}(k,\, N_{t},\, \sigma,\, \mu) &=& \frac{8 \pi^2}{3} \, {\mathcal A}_{{\mathcal R}} \, \, \epsilon \,\, {\mathcal Q}_{E}(\sigma,\,\mu) \biggl(\frac{k}{a H}\biggr)^{7 - 2 \sigma} \,\, e^{ 2 \mu N_{t} ( \sigma  -2)}.
\label{om3a}
\end{eqnarray}
Using Eq. (\ref{slope}) into Eqs. (\ref{om2a}) and (\ref{om3a}), the magnetic and electric spectral indices are, respectively, $n_{B} = 6 - 2 \sigma$ and $n_{E} = 8 - 2 \sigma$; moreover, since $\sigma >1$ the magnetic spectral index is bounded from above, i.e. $n_{B} < 4$. Eliminating $\sigma$ between the explicit expressions of $n_{B}$ and $n_{E}$, the power spectra of Eqs. (\ref{om2a}) and (\ref{om3a}) 
are phrased  in terms of $n_{B}$ and $\mu$:
\begin{eqnarray} 
\Omega_{B}(k,\, N_{t},\, n_{B},\, \mu) &=& \frac{8 \pi^2}{3} \, {\mathcal A}_{{\mathcal R}} \, \, \epsilon \,\, {\mathcal Q}_{B}(n_{B},\,\mu) \biggl(\frac{k}{a H}\biggr)^{n_{B} -1} \,\, e^{\mu N_{t} (4 - n_{B})},
\label{om4a}\\
\Omega_{E}(k,\, N_{t},\, n_{B},\, \mu) &=& \frac{8 \pi^2}{3} \, {\mathcal A}_{{\mathcal R}} \, \, \epsilon \,\, {\mathcal Q}_{E}(n_{B},\,\mu) \biggl(\frac{k}{a H}\biggr)^{n_{B} +1} \,\, e^{\mu N_{t} ( 2 - n_{B})},
\label{om5a}
\end{eqnarray}
where  the prefactors  ${\mathcal Q}_{B}(n_{B},\,\mu)$ and ${\mathcal Q}_{E}(n_{B},\,\mu)$ are, in this case:
\begin{equation}
{\mathcal Q}_{B}(n_{B},\,\mu) = \frac{2^{3 - n_{B}}}{\pi^3} \, \Gamma^2\biggl(\frac{6 - n_{B}}{2}\biggr) | 1 + \mu|^{5 - n_{B}}, \qquad \frac{{\mathcal Q}_{B}(n_{B},\,\mu)}{{\mathcal Q}_{E}(n_{B},\,\mu)}=
( 4 - n_{B})^2 |1 +\mu|^2.
\label{def6a}
\end{equation}
If  $n_{B} \to 1$  in Eqs. (\ref{om4a}) and (\ref{om5a}) the magnetic power spectrum is scale-invariant while the electric power spectrum is blue, i.e.
\begin{eqnarray}
\Omega_{B}(k,\, N_{t},\, 1,\, \mu) &=& \frac{8 \pi^2}{3} \, {\mathcal A}_{{\mathcal R}} \, \, \epsilon \,\, {\mathcal Q}_{B}(1,\,\mu)  \,\, e^{3 \, \mu\, N_{t}},
\label{sia1}\\
\Omega_{E}(k,\, N_{t},\, 1,\, \mu) &=& \frac{8 \pi^2}{3} \, {\mathcal A}_{{\mathcal R}} \, \, \epsilon \,\, {\mathcal Q}_{E}(1,\,\mu) \biggl(\frac{k}{a H}\biggr)^{2} \,\, e^{\mu N_{t}}.
\label{sia2}
\end{eqnarray}
If $n_{B} \to -1$ the electric power spectrum is scale-invariant while the magnetic power spectrum is sharply red:
\begin{eqnarray}
\Omega_{B}(k,\, N_{t},\, -1,\, \mu) &=& \frac{8 \pi^2}{3} \, {\mathcal A}_{{\mathcal R}} \, \, \epsilon \,\, {\mathcal Q}_{B}(-1,\,\mu) \biggl(\frac{k}{a H}\biggr)^{-2} \,\, e^{ 5 \mu N_{t}},
\label{sib1}\\
\Omega_{E}(k,\, N_{t},\, -1,\, \mu) &=& \frac{8 \pi^2}{3} \, {\mathcal A}_{{\mathcal R}} \, \, \epsilon \,\, {\mathcal Q}_{E}(-1,\,\mu)  \,\, e^{3 \mu N_{t} }.
\label{sib2}
\end{eqnarray}
Recalling  Eq. (\ref{fsol}), the relation among $\sigma$, $\mu$ and $\nu$ is given by $\sigma= \nu/(1+ \mu)$. Consequently,  in the limit $\mu \to 0$ and $n_{B} \sim 1$ we also have $\sigma = \nu = 5/2$. In the latter case the magnetic power spectrum
at the time of gravitational collapse can be estimated as\footnote{We express the fields in Gauss and  $1\, \mathrm{nG} = 10^{-9} \,\, \mathrm{G}$.}  $\sqrt{P_{B}} \simeq {\mathcal O}(0.01\, \mathrm{nG})$. This is the result  found in \cite{DT5,DT9,duality3} and it is compatible 
with the origin of large-scale magnetic fields. 

The magnetogenesis requirements \cite{mgenesis,DT5,DT9,duality3} 
 roughly demand that the magnetic fields at the time of the gravitational collapse of the protogalaxy should be approximately larger than a (minimal) field which can be estimated between $10^{-16}$ nG and $10^{-11}$ nG. The most optimistic estimate is derived by assuming 
that every rotation of the galaxy would increase the magnetic field of one efold. The number of galactic rotations since the 
collapse of the protogalaxy can be estimated between $30$ and $35$, 
leading approximately to  a purported growth of $13$ orders of magnitude. During collapse of the protogalaxy compressional amplification will 
increase the field of about $5$ orders of magnitude. Thus the required seed field at the onset of the gravitational collapse must be, at least, as large as $10^{-15}$ nG or, more realistically, larger than $10^{-11}$ nG \cite{DT5,duality3}.  For $\sigma >1$, 
${\mathcal A}_{{\mathcal R}}= 2.41\times 10^{-9}$ and $\epsilon=0.01$ the magnetic power spectrum at the onset of the 
gravitational collapse of the protogalaxy can be written as:
\begin{equation}
\frac{P_{B}(k, N_{t}, n_{B}, \mu)}{\mathrm{G}^2} =  10^{-21.05} \biggl(\frac{k}{H_{0}}\biggr)^{n_{B} -1} e^{ - (n_{B}-1) N_{\mathrm{max}}} \, e^{\mu N_{t} ( 4 - n_{B})}.
\label{inc3}
\end{equation}
Consider now the case when both spectra are strongly increasing or, as we say for short, violet. 
For instance, if $n_{B} = 3$ the magnetic and the electric spectra are given, respectively, by: 
\begin{eqnarray} 
\Omega_{B}(k,\, N_{t},\, 3,\, \mu) &=& \frac{8 \pi^2}{3} \, {\mathcal A}_{{\mathcal R}} \, \, \epsilon \,\, {\mathcal Q}_{B}(3,\,\mu) \biggl(\frac{k}{a H}\biggr)^{2} \,\, e^{ \mu N_{t} },
\label{inc1}\\
\Omega_{E}(k,\, N_{t},\, 3, \, \mu) &=& \frac{8 \pi^2}{3} \, {\mathcal A}_{{\mathcal R}} \, \, \epsilon \,\, {\mathcal Q}_{E}(3,\,\mu) \biggl(\frac{k}{a H}\biggr)^{4} \,\, e^{- \mu N_{t} }.
\label{inc2}
\end{eqnarray}
The constraints on the violet and blue spectra are imposed at $x \simeq 1$;  these scales are actually washed out by the finite value of the conductivity and, in this sense, 
this requirement is rather conservative.  The requirements  $\Omega_{B}(a H,\, N_{t},\, 3,\, \mu) < 10^{-3}$ and  $\Omega_{E}(a H,\, N_{t},\, 3,\, \mu) < 10^{-3}$ cannot be jointly satisfied for $n_{B} =3$ as it is clear from Eqs. (\ref{inc1})--(\ref{inc2}). 
The same conclusion holding for $n_{B} =3$ can be extended to the case $n\geq 2$; from Eqs. (\ref{om4a})--(\ref{om5a})  the following conditions can be derived for $k \simeq a H$:
\begin{eqnarray}
\Omega_{B}(a H,\, N_{t},\, n_{B},\, \mu) &=& \frac{8 \pi^2}{3} \, {\mathcal A}_{{\mathcal R}} \, \, \epsilon \,\, {\mathcal Q}_{B}(n_{B},\,\mu) \,\, e^{\mu N_{t} (4 - n_{B})} < 10^{-3},
\label{inc3a}\\
\Omega_{E}(a H ,\, N_{t},\, n_{B},\, \mu) &=& \frac{8 \pi^2}{3} \, {\mathcal A}_{{\mathcal R}} \, \, \epsilon \,\, {\mathcal Q}_{E}(n_{B},\,\mu) \,\, e^{\mu N_{t} ( 2 - n_{B})}< 10^{-3},
\label{inc4a}
\end{eqnarray}
cannot be jointly satisfied. The conditions imposed by Eqs. (\ref{inc3a})--(\ref{inc4a}) can be relaxed if the maximal wavenumber 
is not given by $ x \sim 1$ but rather by $x_{\sigma} = k_{\sigma}/(a H) \simeq 10^{-13}$. In the latter case larger spectral indices $n_{B} \geq  2$ can be accommodated and the parameter space may get even wider. In what follows this potentially interesting aspect shall be neglected.
\begin{figure}[!ht]
\centering
\includegraphics[height=7.5cm]{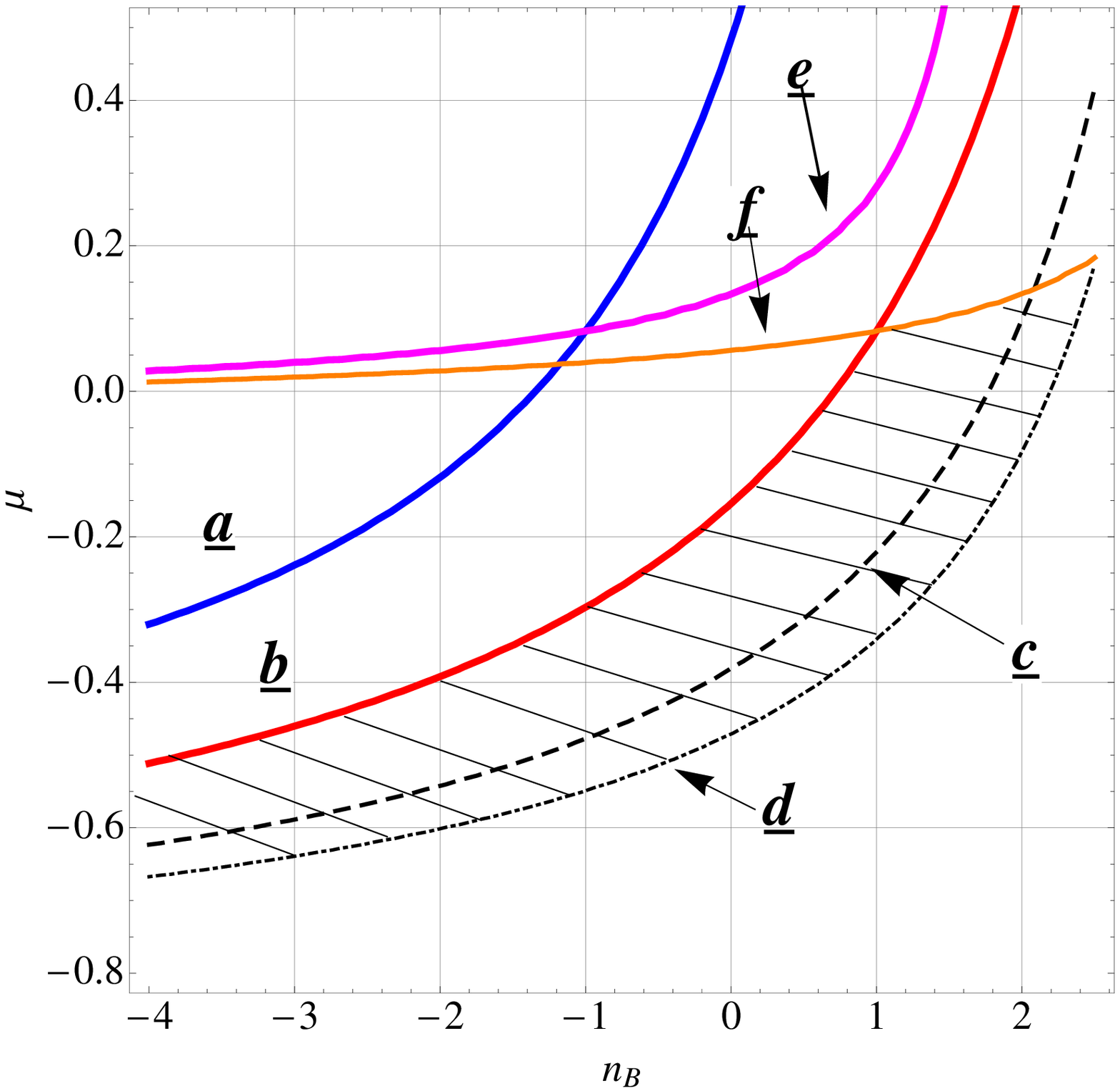}
\includegraphics[height=7.5cm]{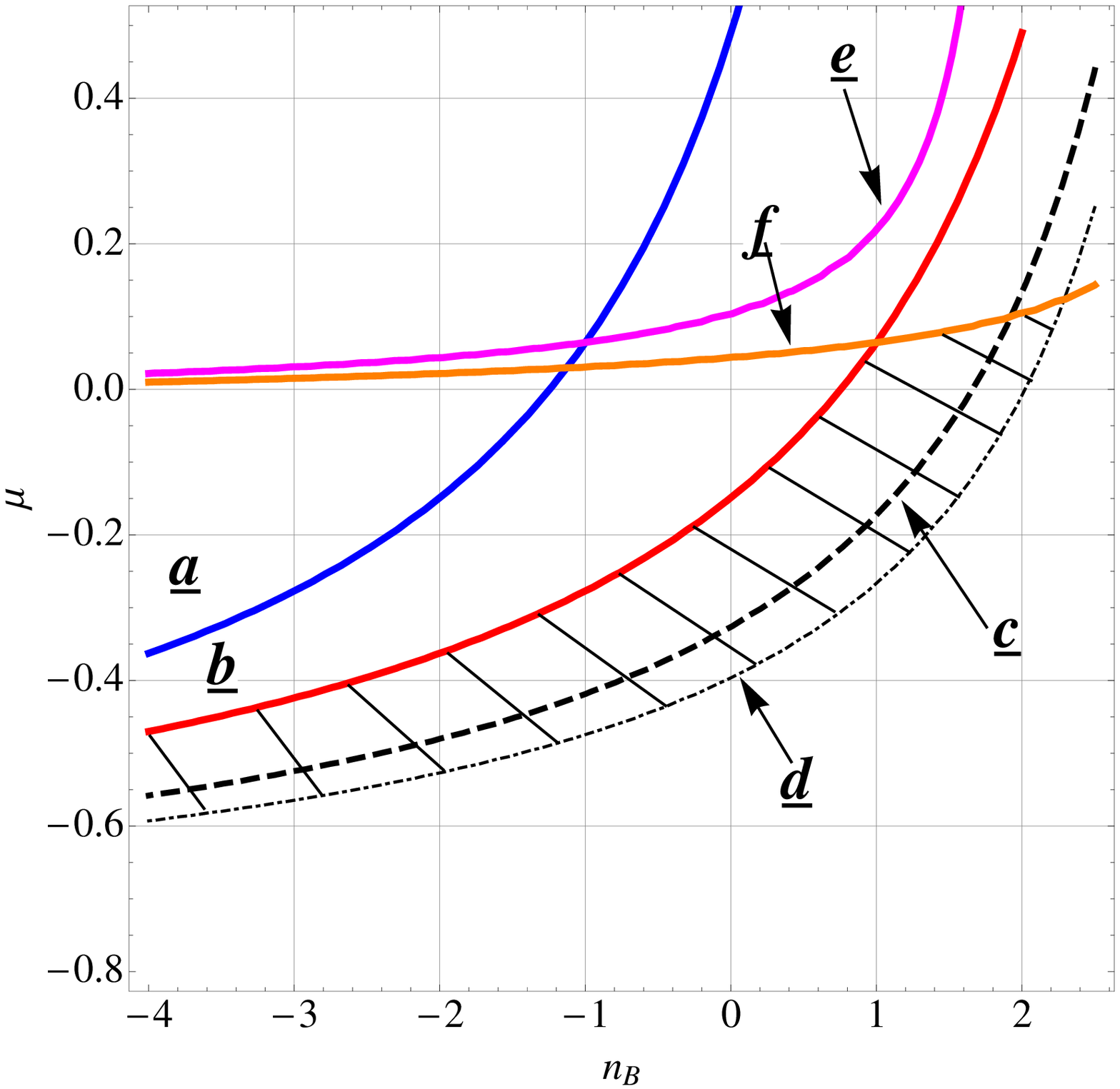}
\caption[a]{The various constraints of the case $\sigma >1$ are illustrated. The shaded area represents the allowed region in the parameter space where the back-reaction constraints are avoided and the minimal magnetogenesis requirements satisfied.}
\label{Figure1}      
\end{figure}

Recalling Eq. (\ref{xx}), Eqs. (\ref{om2a})--(\ref{om3a}) can be directly expressed in terms of $N_{\mathrm{max}}$ and $N_{t}$:
\begin{eqnarray} 
&&\Omega_{B}(k,\, N_{t},\, n_{B},\, \mu) = \frac{8 \pi^2}{3} {\mathcal A}_{{\mathcal R}}\,  \epsilon\, {\mathcal Q}_{B}(n_{B},\,\mu) \biggl(\frac{k}{H_{0}}\biggr)^{n_{B} -1} e^{{\mathcal F}_{B}(\mu, N_{t},n_{B}) },
\label{om7a}\\
&&\Omega_{E}(k,\, N_{t},\, n_{B},\, \mu) = \frac{8 \pi^2}{3}  {\mathcal A}_{{\mathcal R}}\, \epsilon \, {\mathcal Q}_{E}(n_{B},\,\mu) \biggl(\frac{k}{H_{0}}\, \biggr)^{n_{B} +1} e^{{\mathcal F}_{E}(\mu, N_{t},n_{B}) }.
\label{om8a}
\end{eqnarray}
where
\begin{eqnarray}
{\mathcal F}_{B}(\mu, N_{t}, N_{\mathrm{max}}, n_{B}) &=& - N_{\mathrm{max}}(n_{B} -1) + \mu N_{t} (4 - n_{B}),
\nonumber\\
{\mathcal F}_{E}(\mu, N_{t}, N_{\mathrm{max}}, n_{B}) &=& - N_{\mathrm{max}}(n_{B} + 1) + \mu N_{t}(2 - n_{B}).
\end{eqnarray}
In Fig. \ref{Figure1}, for two illustrative choices of the parameters, we plot six different contours 
corresponding to the curves labeled by  $(a)$, $(b)$, $(c)$, $(d)$, $(e)$ and $(f)$:
\begin{itemize} 
\item{} the curves $(a)$ and $(b)$ correspond, respectively, to  
to $\Omega_{E}(k,N_{t}, n_{B}, \mu) = 10^{-3}$ and  $\Omega_{B}(k,N_{t}, n_{B}, \mu) = 10^{-3}$ when 
$k= 1\, \mathrm{Mpc}^{-1}$ and  $N_{t} = N_{\mathrm{max}}$ (plot on the left) and when $k= 10^{-4}\, \mathrm{Mpc}^{-1}$ and $N_{t} = 80 > N_{\mathrm{max}}$ (plot on the right); these low-frequency bounds are the most constraining for red spectra;
\item{} the curves $(c)$ and $(d)$ correspond, respectively,  to  $P_{B}(k,N_{t}, n_{B}, \mu) = 10^{-22} \, \mathrm{nG}^2$ and to 
$P_{B}(k,N_{t}, n_{B}, \mu) = 10^{-32} \, \mathrm{nG}^2$; the parameters of the plots are  $k= 1\, \mathrm{Mpc}^{-1}$ (roughly corresponding to the scale of protogalactic collapse) for both plots; moreover the total number of efolds is such that  $N_{t} = N_{\mathrm{max}}$ (plot on the left) and  $N_{t} = 80 > N_{\mathrm{max}}$ (plot on the right);
\item{} the curves $(f)$ and $(e)$ illustrate, respectively,  the contours of Eqs. (\ref{inc3a}) and (\ref{inc4a}) for $N_{t} = N_{\mathrm{max}}$ (plot on the left) and for  $k= 10^{-4}\, \mathrm{Mpc}^{-1}$ and $N_{t} = 80 > N_{\mathrm{max}}$ (plot on the right); these requirements are the most 
constraining for blue and violet spectra;
\item{} the shaded area is the allowed region in the parameter space
where  the back reaction constraints are safely enforced and the magnetogenesis requirements are met. 
\end{itemize}
Note that 
when $N_{t}$ increases beyond $N_{\mathrm{max}}$ the area of the allowed region gets narrower. 

The shaded area of Fig. \ref{Figure1} can be compared with Fig. \ref{Figure0}. For $\sigma >1$ the relation to the magnetic power spectrum is given by $ \sigma = (6 - n_{B})/2$. The conditions 
implied by Fig. \ref{Figure0} demand that $\chi_{E}$ and $\chi_{B}$ are both decreasing provided 
the two following inequalities are simultaneously satisfied:
\begin{equation}
n_{B} \geq \frac{6 \mu + 5}{\mu + 1}, \qquad n_{B} \geq \frac{4 \mu + 5}{\mu + 1},
\label{in1}
\end{equation}   
where the first inequality refers to $\chi_{E}$ while the second inequality refers to $\chi_{B}$.  
Since $n_{B}$ is bounded from above (i.e. $n_{B} < 4$ because $\sigma > 1$) it follows that for $\mu < -1$ the first inequality of Eq. (\ref{in1}) 
is never satisfied while the second may or may not be satisfied. Thus, for $\mu <-1$ and $n_{B} < 4$, 
$\chi_{E}$ must necessarily increase while $\chi_{B}$ may either increase or decrease. 

If $\mu >-1$ the second inequality of Eq. (\ref{in1}) is always verified since $n_{B} \to 4$ is the asymptote of the corresponding hyperbola. 
The first inequality may or may not be satisfied. Moreover, since the line $n_{B} =4$ intersects the hyperbola, $\mu$ will be bounded 
from below by the asymptote and from above by the intersection; we will then have that the relevant range is $-1 < \mu \leq -1/2$. 
We can then say that for  $-1 < \mu \leq -1/2$ and $n_{B}<4$ the magnetic susceptibility $\chi_{B}$ is always decreasing while $\chi_{E}$ may either increase or decrease.

So we can conclude by saying that magnetogenesis is viable and the back reaction constraints safely satisfied 
in the regions illustrated in Fig. \ref{Figure2}. The models are dynamically realized in a number of ways 
but, in this case (i.e. $\sigma >1$) {\em at least one of the susceptibilities must be increasing}.

\subsection{The case $0 < \sigma < 1$}
In the remaining two regions of the parameter space the analysis follows the same steps already outlined in the 
case $\sigma > 1$. The logic of the discussion will be exactly the same so that we shall skip the details and stick to the results. 
If $0<\sigma <1$ the power spectra of Eqs. (\ref{om2}) and (\ref{om3}) become:
\begin{eqnarray} 
\Omega_{B}(k,\, N,\, \sigma,\, \mu) &=& \frac{8 \pi^2}{3} \, {\mathcal A}_{{\mathcal R}} \, \, \epsilon \,\, {\mathcal Q}_{B}(\sigma,\,\mu) \biggl(\frac{k}{a H}\biggr)^{5 - 2 \sigma} \,\, e^{ 2 \mu N_{t} ( \sigma  -1)},
\label{om2b}\\
\Omega_{E}(k,\, N,\, \sigma,\, \mu) &=& \frac{8 \pi^2}{3} \, {\mathcal A}_{{\mathcal R}} \, \, \epsilon \,\, {\mathcal Q}_{E}(\sigma,\,\mu) \biggl(\frac{k}{a H}\biggr)^{3 + 2 \sigma} \,\, 
e^{ -2 \mu N_{t} \sigma }.
\label{om3b}
\end{eqnarray}
\begin{figure}[!ht]
\centering
\includegraphics[height=7.5cm]{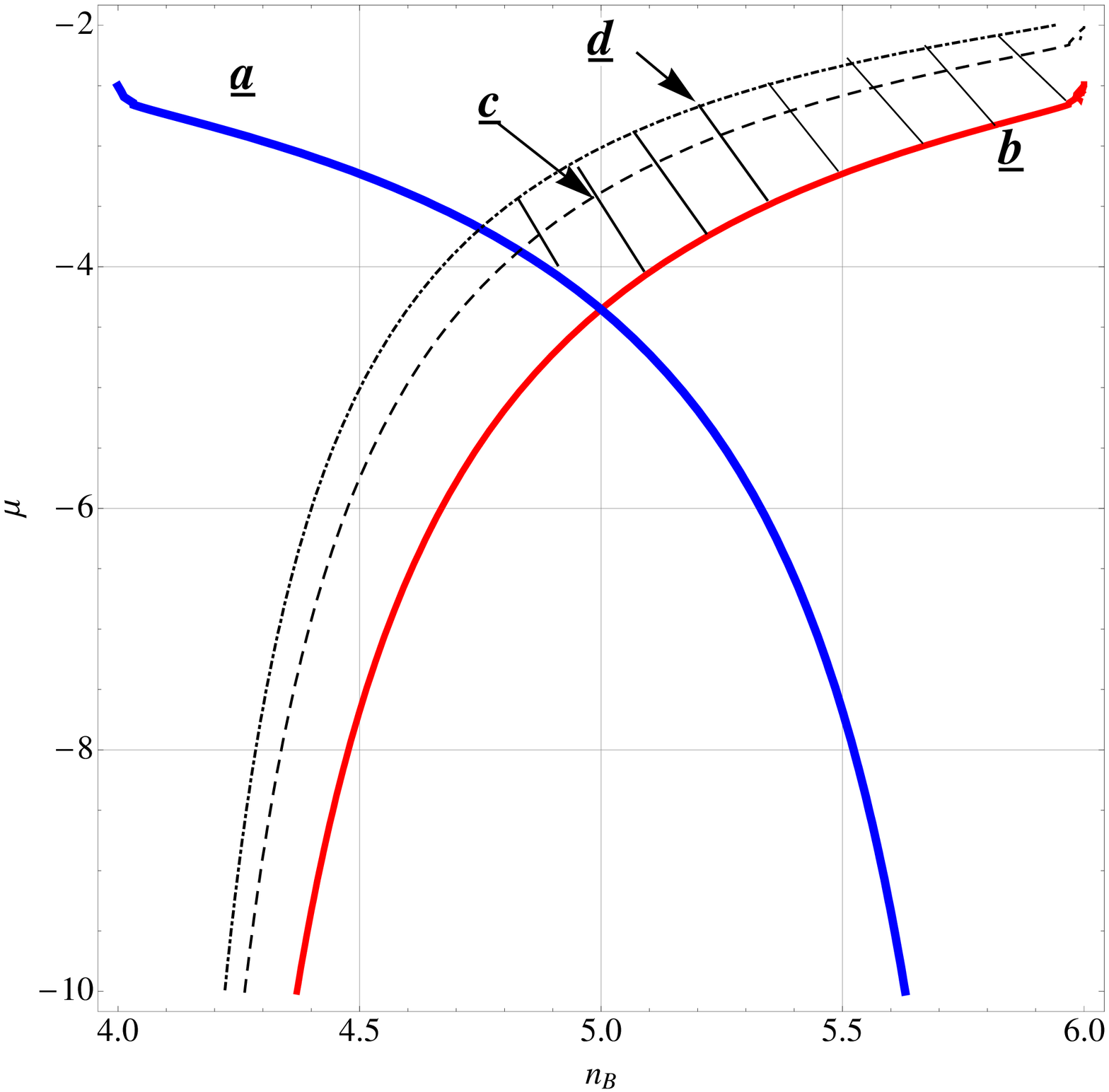}
\includegraphics[height=7.5cm]{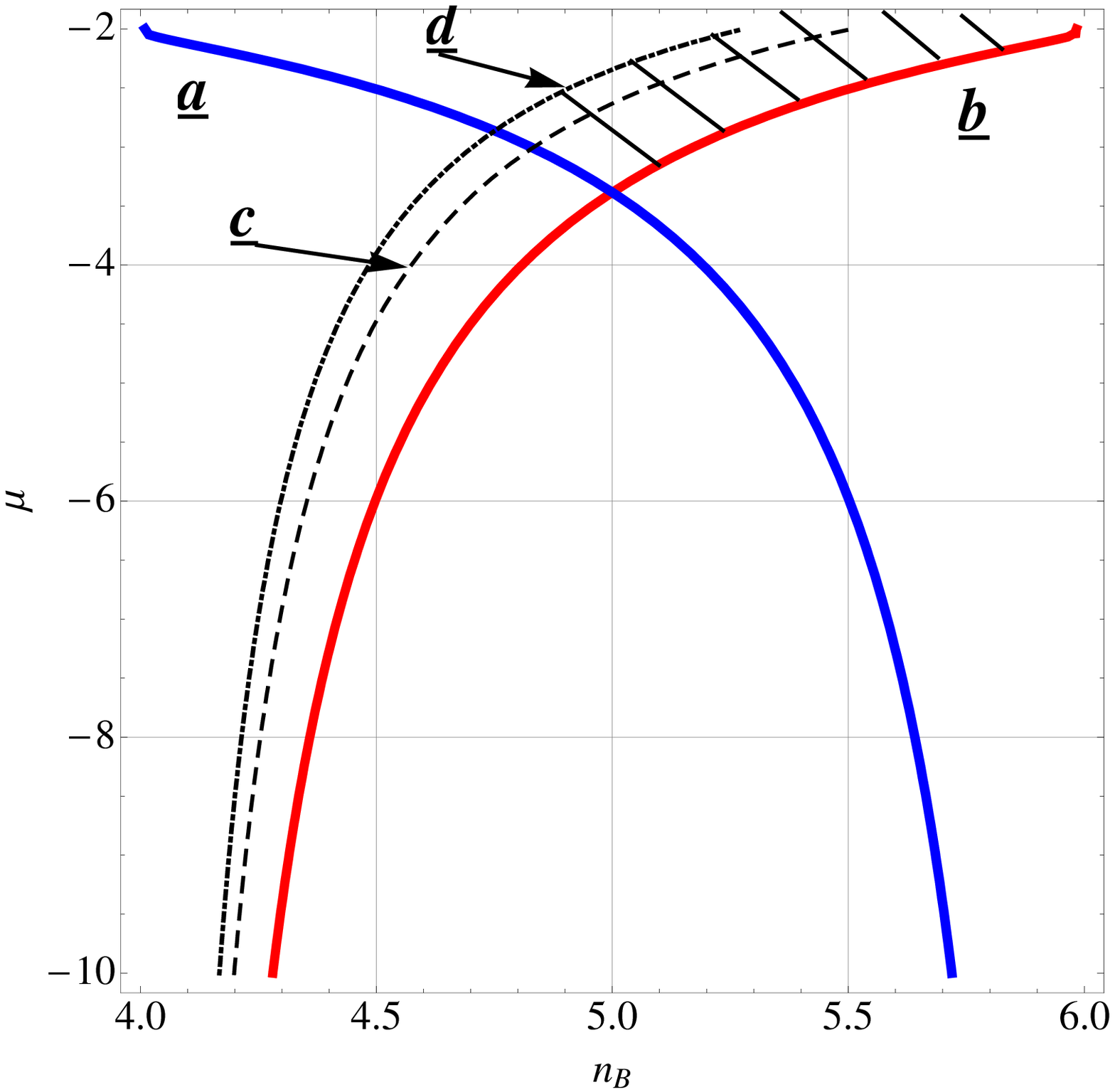}
\caption[a]{The exclusion plot in the case $0< \sigma <1$. The shaded area illustrates the region where the magnetogenesis requirements 
and the large-scale back-reaction constraints are satisfied. Since, in this case, the spectra are always violet the most significant constraints 
arise from the maximally amplified length-scale. These constraints cannot be jointly satisfied within the shaded area of this plot so that 
magnetogenesis is not viable in this case.}
\label{Figure2}      
\end{figure}
From Eq. (\ref{om2b}) $\sigma$ can be expressed in terms of the magnetic spectral index $n_{B}$ as $\sigma = (6 - n_{B})/2$. However,  since $0 < \sigma < 1$ we must 
also demand, this time, that $4 < n_{B}< 6$. In terms of $n_{B}$ Eqs. (\ref{om2b}) and (\ref{om3b}) can be written as:
\begin{eqnarray} 
\Omega_{B}(k,\, N_{t},\, n_{B},\, \mu) &=& \frac{8 \pi^2}{3} \, {\mathcal A}_{{\mathcal R}} \, \, \epsilon \,\, {\mathcal Q}_{B}(n_{B},\,\mu) \biggl(\frac{k}{a H}\biggr)^{n_{B} -1} \,\, e^{\mu N_{t} (4 - n_{B})},
\label{om4b}\\
\Omega_{E}(k,\, N_{t},\, n_{B},\, \mu) &=& \frac{8 \pi^2}{3} \, {\mathcal A}_{{\mathcal R}} \, \, \epsilon \,\, {\mathcal Q}_{E}(n_{B},\,\mu) \biggl(\frac{k}{a H}\biggr)^{9 -n_{B} } \,\, e^{ - \mu N_{t}( 6 - n_{B})},
\label{om5b}
\end{eqnarray}
where, in this case, 
\begin{eqnarray}
{\mathcal Q}_{B}(n_{B},\,\mu) &=& \frac{2^{3 - n_{B}}}{\pi^3} \, \Gamma^2\biggl(\frac{6 - n_{B}}{2}\biggr) | 1 + \mu|^{5 - n_{B}}, \nonumber\\
 {\mathcal Q}_{E}(n_{B},\,\mu) &=& \frac{2^{n_{B}-7}}{\pi^3} \, \Gamma^2\biggl(\frac{n_{B}-4}{2}\biggr) | 1 + \mu|^{n_{B}-5}.
\label{def6b}
\end{eqnarray}
In the range $4 < n_{B} < 6$ the scale-invariant magnetic power spectrum 
and the scale-invariant electric power spectrum are both impossible since the corresponding values of $n_{B}$ are located outside 
the interval. 

 In Fig. \ref{Figure2} the conditions $\Omega_{E}(k,N_{t}, n_{B}, \mu) = 10^{-3}$ and $\Omega_{B}(k,N_{t}, n_{B}, \mu) = 10^{-3}$ correspond, respectively, to the curves $(a)$ and $(b)$ where $k = 1\,\, \mathrm{Mpc}^{-1}$ and $N_{t} = N_{\mathrm{max}}$ (plot on the left); similarly in the plot on the right
$k = 10^{-4}\, \mathrm{Mpc}^{-1}$ and $N_{t} =80$. The curves $(c)$ and $(d)$ denote the same magnetogensis requirements
of Fig. \ref{Figure1} but illustrated in terms of the power spectra (\ref{om4b}) and (\ref{om5b}). The shaded area is the region where 
the (large-scale) back-reaction constraints and the magnetogenesis bounds are jointly satisfied. In spite of that the shaded area must be excluded. Indeed, as it is clear from Eqs. (\ref{om4b}) and (\ref{om5b}), for $4 < n_{B} < 6$ both electric and magnetic spectra are 
violet. Hence the most significant constraints will come from the region $x \sim 1$ (or $k \sim a H$).  Setting $k\sim a\,H$ in Eqs. (\ref{om4b}) and (\ref{om5b}) 
\begin{eqnarray} 
\Omega_{B}(k,\, N_{t},\, n_{B},\, \mu) &=& \frac{8 \pi^2}{3} \, {\mathcal A}_{{\mathcal R}} \, \, \epsilon \,\, {\mathcal Q}_{B}(n_{B},\,\mu) \,\, e^{\mu N_{t} (4 - n_{B})}< 10^{-3},
\label{ccc1}\\
\Omega_{E}(k,\, N_{t},\, n_{B},\, \mu) &=& \frac{8 \pi^2}{3} \, {\mathcal A}_{{\mathcal R}} \, \, \epsilon \,\,  {\mathcal Q}_{B}(n_{B},\,\mu)  \,\, e^{ - \mu N_{t}( 6 - n_{B})}< 10^{-3}.
\label{ccc2}
\end{eqnarray}
These conditions are jointly verified, as it can be easily checked, provided the values of $\mu$  are well {\em above} the shaded area of Fig. \ref{Figure2}.  Since no overlaps between the regions exists there are no viable models of magnetogenesis when $0 < \sigma < 1$.

\subsection{The case $\sigma < 0$}
Inserting Eqs. (\ref{sp1c})--(\ref{sp2c}) into  Eqs. (\ref{om2})--(\ref{om3}),  the explicit form of
the power spectra in the case $\sigma <0$ is:
\begin{eqnarray} 
\Omega_{B}(k,\, N_{t},\, \sigma,\, \mu) &=& \frac{8 \pi^2}{3} \, {\mathcal A}_{{\mathcal R}} \, \, \epsilon \,\, {\mathcal Q}_{B}(\sigma,\,\mu) \biggl(\frac{k}{a H}\biggr)^{5 + 2 \sigma} \,\, e^{-2 \mu N_{t} ( \sigma  +1)},
\label{om2c}\\
\Omega_{E}(k,\, N_{t},\, \sigma,\, \mu) &=& \frac{8 \pi^2}{3} \, {\mathcal A}_{{\mathcal R}} \, \, \epsilon \,\, {\mathcal Q}_{E}(\sigma,\,\mu) \biggl(\frac{k}{a H}\biggr)^{3 + 2 \sigma} \,\, 
e^{-2 \mu N_{t}  \sigma }.
\label{om3c}
\end{eqnarray}
\begin{figure}[!ht]
\centering
\includegraphics[height=7.5cm]{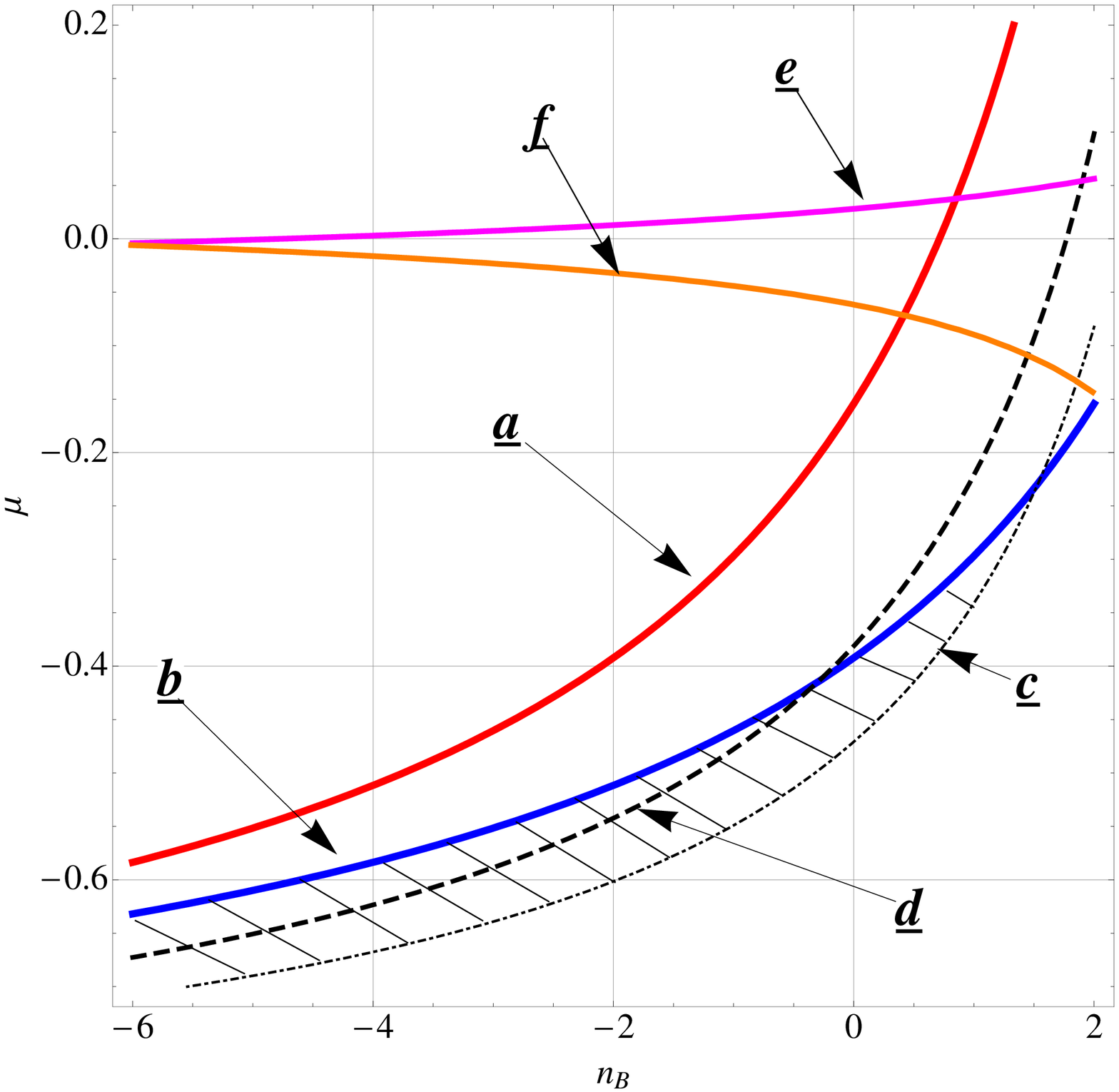}
\includegraphics[height=7.5cm]{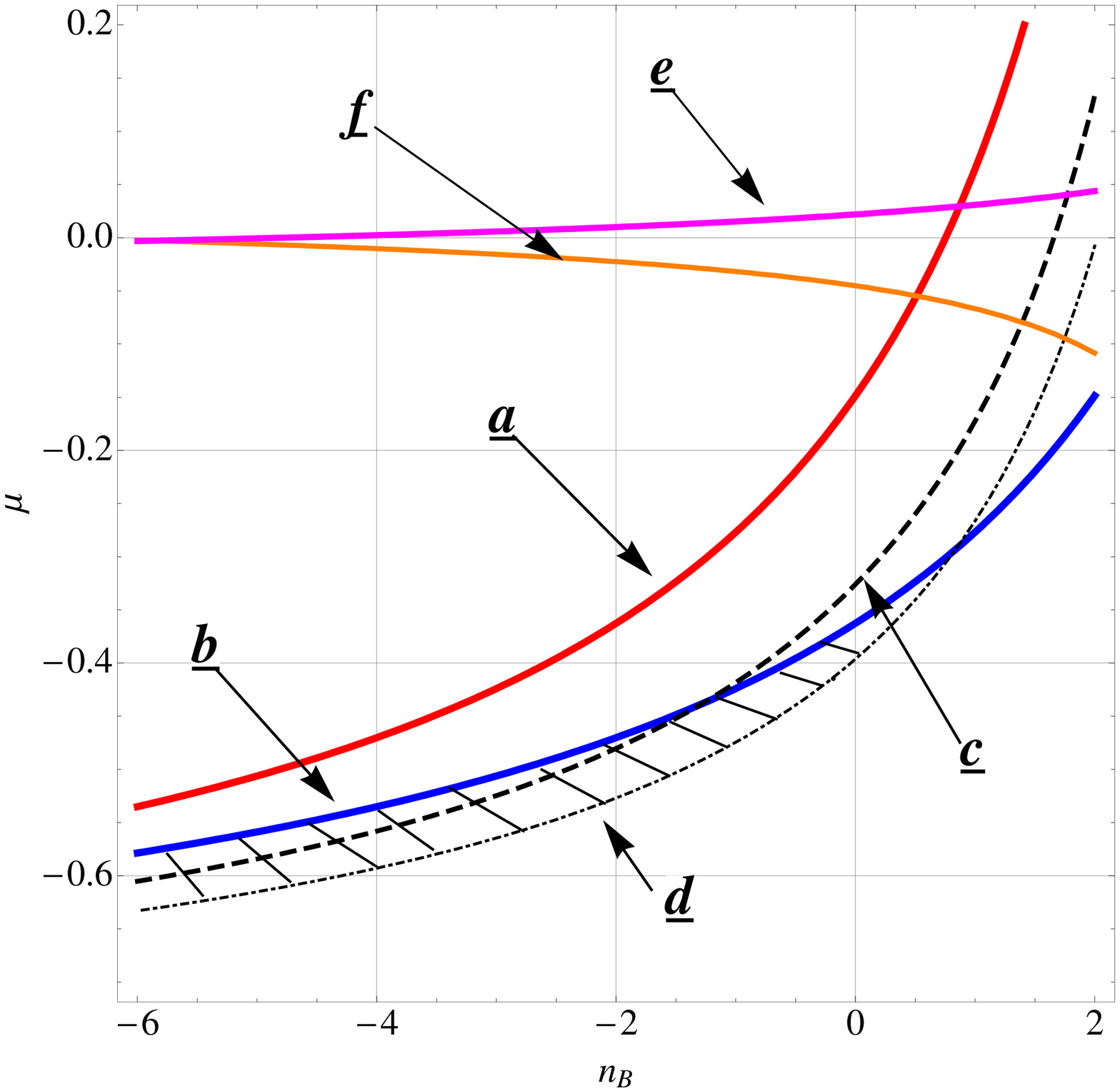}
\caption[a]{The exclusion plot in the case $\sigma <0$. As in Figs. \ref{Figure1} the shaded area illustrates the allowed region in the parameter space where the magnetogenesis requirements are met and the back-reaction constrains satisfied.}
\label{Figure3}      
\end{figure}
According to Eqs. (\ref{om2c}) and (\ref{om3c}) and using Eq. (\ref{slope}) the magnetic and electric spectral indices are, respectively, $n_{B} = 6 + 2 \sigma$ and $n_{E} = 4 + 2 \sigma$. Since $\sigma = (n_{B}-6)/2$, the condition $\sigma <0$ implies  $n_{B} < 6$.  Elimitating $\sigma$ in favour of $n_{B}$, Eqs. (\ref{om2c}) and (\ref{om3c})  become
\begin{eqnarray}
\Omega_{B}(k,\, N_{t},\, n_{B},\, \mu) &=& \frac{8 \pi^2}{3} \, {\mathcal A}_{{\mathcal R}} \, \, \epsilon \,\, {\mathcal Q}_{B}(n_{B},\,\mu) \biggl(\frac{k}{a H}\biggr)^{n_{B}-1} \,\, e^{- \mu N_{t} ( n_{B} -4)},
\label{om4c}\\
\Omega_{E}(k,\, N_{t},\, n_{B},\, \mu) &=& \frac{8 \pi^2}{3} \, {\mathcal A}_{{\mathcal R}} \, \, \epsilon \,\, {\mathcal Q}_{E}(n_{B},\,\mu) \biggl(\frac{k}{a H}\biggr)^{n_{B}-3} \,\, 
e^{- \mu N_{t} (n_{B}-6)},
\label{om5c}
\end{eqnarray}
where
\begin{eqnarray}
&&{\mathcal Q}_{B}(n_{B},\,\mu) = \frac{2^{3 - n_{B}}}{\pi^3} \, \Gamma^2\biggl(\frac{6 - n_{B}}{2}\biggr) | 1 + \mu|^{5 - n_{B}}, 
\nonumber\\
&& {\mathcal Q}_{E}(n_{B},\,\mu) = \frac{2^{5-n_{B}}}{\pi^3} \, \Gamma^2\biggl(\frac{8-n_{B}}{2}\biggr) | 1 + \mu|^{7-n_{B}}, 
\label{def6c}
\end{eqnarray}
The scale-invariant magnetic power spectrum occurs for $n_{B} =1$:
\begin{eqnarray}
\Omega_{B}(k,\, N_{t},\, 1,\, \mu) &=& \frac{8 \pi^2}{3} \, {\mathcal A}_{{\mathcal R}} \, \, \epsilon \,\, {\mathcal Q}_{B}(1,\,\mu) \,\, e^{3 \mu N_{t}},
\label{om4d}\\
\Omega_{E}(k,\, N_{t},\, 1,\, \mu) &=& \frac{8 \pi^2}{3} \, {\mathcal A}_{{\mathcal R}} \, \, \epsilon \,\, {\mathcal Q}_{E}(1,\,\mu) \biggl(\frac{k}{a H}\biggr)^{-2} \,\, 
e^{5 \mu N_{t}}.
\label{om5d}
\end{eqnarray}
The scale-invariant electric power spectrum occurs for $n_{B} =3$:
\begin{eqnarray}
\Omega_{B}(k,\, N_{t},\, 3,\, \mu) &=& \frac{8 \pi^2}{3} \, {\mathcal A}_{{\mathcal R}} \, \, \epsilon \,\, {\mathcal Q}_{B}(3,\,\mu) \,\, \biggl(\frac{k}{a H}\biggr)^{2}  e^{ \mu N_{t}},
\label{om4e}\\
\Omega_{E}(k,\, N_{t},\, 3,\, \mu) &=& \frac{8 \pi^2}{3} \, {\mathcal A}_{{\mathcal R}} \, \, \epsilon \,\, {\mathcal Q}_{E}(3,\,\mu) \,\, 
e^{3 \mu N_{t}}.
\label{om5e}
\end{eqnarray}
By looking at Eqs. (\ref{om4d})--(\ref{om5d}) and (\ref{om4e})--(\ref{om5e}) it can be argued that $\mu$ must be negative to have compatibility 
of the spectra with the critical density bound. This conclusion is corroborated by the exclusion 
plot in the $(\mu,\, n_{B})$ plane which is illustrated in Fig. \ref{Figure3} where  the shaded area represents the allowed region in the parameter space where the magnetogenesis requirements are met and the back-reaction constrains satisfied both at large and small scales. 
The various labels on the curves have the same meaning 
of the ones already discussed in connection with Fig. \ref{Figure1}.
 
The results of Fig. \ref{Figure3} can be considered in conjunction with 
 the ones of Fig. \ref{Figure0}. If $\sigma<0$  the relation of $\sigma$ and $n_{B}$  implies that the electric and magnetic susceptibilities are both decreasing provided the following pair  of inequalities is satisfied:
 \begin{equation}
 n_{B} < \frac{8 \mu  + 7}{\mu +1}, \qquad n_{B} < \frac{6 \mu  + 7}{\mu +1},
 \label{ineq2}
 \end{equation}
 where first inequality refers to $\chi_{B}$ while the second to $\chi_{E}$. 
 If $\mu < -1$ the first inequality is always verified since $n_{B} < 6$ and the asymptote 
 of the first hyperbola is $n_{B} =8$; the second inequality may or may not be verified. Therefore, for $\mu < -1$ and $n_{B} < 6$,
  $\chi_{B}$  is always decreasing while $\chi_{E}$ may either increase or decrease. 
If $\mu > -1$ we have somehow an opposite situation so that the second inequality of Eq. (\ref{ineq2}) is always 
verified while the first inequality may or may not be verified; furthermore, since   $n_{B} =6$ intersects the first hyeprbola in $\mu = -1/2$ we 
must have $-1 < \mu < -1/2$. This means that $\chi_{E}$ is always decreasing while $\chi_{B}$ may or may not decrease.

We can therefore summarize by saying that in the shaded area of  Fig. \ref{Figure3} it is  possible to find viable models 
of magnetogensis in two complementary cases, i.e. {\em either when the susceptibilities are both 
decreasing} during the quasi-de Sitter stage {\em or when one of the susceptibilities increases and the other decreases}.

\subsection{Side remarks and specific cases}
 The borderline situations $\sigma=0$ and $\sigma=1$ must be separately discussed. If $\sigma =0$ the power spectra 
of Eqs. (\ref{sp1c}) and (\ref{sp2c}) are, up to logarithmic corrections,  $P_{B} \propto H^4 x^{5} y^{2\mu}$ and $P_{E} \propto H^4 x^{3}$.  If $\sigma=1$ we have, by duality, $P_{B}(x,\, y,\, \mu,\, 1) = P_{E}(x,\, y,\, \mu,\, 0)$ and $P_{E}(x,\, y,\, \mu,\, 1) = P_{B}(x,\, y,\, \mu,\, 0)$. None 
of these two cases is particularly relevant from the phenomenological viewpoint.

In Fig. \ref{Figure1} the region of the parameter space where $\mu \to 0$ is allowed: 
whenever $\mu \to 0$ there is a region in the parameter 
space where the two susceptibilities coincide, the back-reaction constraints are avoided 
and the magnetogenesis constraints satisfied. This is consistent with earlier results (see, e.g. \cite{DT9,duality3}).
The same exercise can be done in the case of Fig. \ref{Figure3} where 
 the situation is different since the region of the parameter space with $\mu =0$ is not included 
in the allowed region of the parameter space. This is a further evidence that the parameter space of the model is wider when the two susceptibilities do not coincide.

\renewcommand{\theequation}{6.\arabic{equation}}
\setcounter{equation}{0}
\section{Concluding remarks}
\label{sec6}
In this paper we investigated the possibility that the  
electric and the magnetic susceptibilities do not coincide during a phase of quasi-de Sitter expansion. 
Using a generalized duality symmetry it is possible to relate the electric and the 
magnetic power spectra of the quantum fluctuations. The parameter space of inflationary magnetogenesis is widened in comparison with the conventional 
situation where the susceptibilities are equal. The minimal magnetogenesis requirements are met in various regions of the parameter space where back-reaction effects are absent. The magnetic fields can be as large as ${\mathcal O}(0.01)$ nG for typical scales ${\mathcal O}(\mathrm{Mpc})$. 
Both strongly coupled and weakly coupled initial conditions are possible but with different spectral features. 

\end{document}